\documentclass[%
preprint,
 amsmath,amssymb,
]{revtex4-1}

\usepackage{graphicx}
\usepackage{bm}
\usepackage[most]{tcolorbox}
\usepackage[english]{babel}
\usepackage[utf8x]{inputenc}
\usepackage[T1]{fontenc}
\usepackage[linewidth=1pt]{mdframed}
\usepackage{lipsum}
\usepackage{capt-of}
\usepackage{soul,xcolor}
\usepackage[normalem]{ulem}

\usepackage[a4paper,top=3cm,bottom=2cm,left=3cm,right=3cm,marginparwidth=1.75cm]{geometry}

\usepackage{amsmath}
\usepackage{graphicx}
\usepackage[colorinlistoftodos]{todonotes}
\usepackage[colorlinks=true, allcolors=blue]{hyperref}
\begin{document}
\newcommand{\affilITAMP}{ITAMP, Harvard-Smithsonian Center for Astrophysics, Cambridge, MA 02138, USA}
\newcommand{\affilNavy}{Department of Physics, The United States Naval Academy, Annapolis, MD 21402, USA}
\newcommand{\affilOU}{Homer L. Dodge Department of Physics and Astronomy, The University of Oklahoma, 440 W. Brooks St.
Norman, OK 73019, USA}

\title{Ultracold Rydberg molecules}
\author{J. P. Shaffer}
\affiliation{\affilOU}
\author{S. T. Rittenhouse}
\affiliation{\affilNavy}
\author{H. R. Sadeghpour}
\affiliation{\affilITAMP}

\email[]{shaffer@nhn.ou.edu}

\date{\today}

\begin{abstract}
Ultracold Rydberg molecules have been extensively studied both theoretically and experimentally. Here the authors review the recent realizations of various ultralong-range Rydberg molecules and macrodimers, and explore their potential for current and future applications in spectroscopy, few- and many-body interactions and quantum information processing.

\end{abstract}
\maketitle

\section*{Introduction}
The spectroscopy of excited atoms was instrumental in the development of early atomic physics as well as the evolution of quantum physics. The Balmer series in hydrogen \cite{Balmer1885}, and the spectra of other atoms, notably the alkali metal atoms by Rydberg \cite{Rydberg1889}, provided some of the most important clues to understanding nature on a microscopic scale \cite{gallagher2005rydberg}.

The modern picture of the exotic molecules described in this brief review emerged from the investigation of spectral line broadening of highly excited atoms.  What happens to an atom with a weakly bound electron in the presence of surrounding atoms or molecules? The answer to this question leads to an intriguing line of inquiry that culminates in our current understanding of ultracold Rydberg molecules, including ultralong-range Rydberg molecules and Rydberg macrodimers, where, respectively, single and double Rydberg excitations are involved in their formation.

The review is organized to first provide a historical footing for much of what will follow. This  perspective is additionally relevant for a proper accounting of how two of the most important topics in low-energy physics, the scattering length and the pseudopotential, emerged. We then describe the ultralong-range Rydberg molecule and Rydberg macrodimer Hamiltonians. Finally, we report on the spectroscopic properties and laboratory realizations of both types of Rydberg molecules. The few-body and many-body implications of these molecules in atomic quantum gases are then discussed.

\section*{Formation of Ultracold Rydberg and Macrodimer Molecules}
\subsection*{Collisional interactions between a Rydberg atom and a ground-state atom}\label{sec:pseudopot}

The first measurements of the collisional properties of Rydberg atoms were made by Amaldi and Segre \cite{amaldi1934}. In their ground-breaking experiment, they observed that the shift of atomic Rydberg lines depended on the type of gas in which the sodium or potassium vapors they studied were immersed. They discovered that  in different gases (H$_2$, N$_2$, He and Ar), the alkali metal Rydberg lines sometimes shifted to the red, and at other times shifted to the blue of the unperturbed atomic lines observed at relatively low gas densities.  It was left to Fermi to develop the first theory of collisional shift and broadening of Rydberg lines. In the process of explaining the experimental observations of Amaldi and Segre, Fermi introduced two important concepts to low-energy physics: the scattering length and the pseudopotential \cite{fermi1934}. 

To appreciate Fermi's insight, consider how a weakly bound electron in a gas interacts with nearby atoms or molecules. If the neighboring species which interact with the Rydberg electron do not possess a permanent electric dipole moment, then the interaction of the electron with the surrounding ground-state atoms is in the first order, the polarization potential. An immediate expectation is that all spectral lines shift to the red, see {\bf Box 1}. 

Fermi recognized that scattering mattered: In the low-energy scattering of the Rydberg electron from  perturber atoms, the wave function will be slowly varying, except near the perturbers. He therefore invented another wave function ${\bar \psi}$ as the mean value of $\psi$ around the perturbers, whose extent was assumed to be small compared with the Rydberg electron de Broglie wavelength, but still large enough to contain a sizable number of perturber atoms. The wave function, regular at the origin, can be written $u(r) = r\psi(r) = (r - a_s(0)){\bar\psi}$, with 
$a_s (0)= -\lim_{k\rightarrow 0} \frac{1}{k} \tan{(\delta_s (k))}$ the $s$-wave electron scattering length, obtained in the zero-momentum limit of the s-wave phase shift, $\delta_s (k)$. In the above, $r$ is the distance from the Rydberg electron to the perturber atoms.

With this approach, the spectral energy shift in the long-wavelength limit becomes, e. g. Eq. 17 in \cite{fermi1934} in units of cm$^{-1}$
\begin{equation}
\Delta_a = \frac{\hbar a_s(0) \rho}{m_ec},
\label{Fermishift}
\end{equation}
with $m_e$ the mass of the electron. The total energy shift is $\Delta_a + \Delta_\epsilon$, where 
\begin{equation}
\Delta_\epsilon \approx 2.6(\frac{4\pi}{3})^{4/3} [\frac{(\epsilon -1)}{16\pi^2\hbar  c}] e^2 \rho^{1/3},
\end {equation}
in units of cm$^{-1}$, is the shift due to dielectric screening of the Rydberg electron by the gas atoms at density $\rho$ and dielectric constant $\epsilon$, see {\bf Box 1}. In the above equations, $e$ is the elementary charge on the electron, $\hbar$ is the reduced Planck constant, and $c$ is the speed of light. In atomic units, $e=\hbar=1$. The crucial point that unveils itself is that when $a_s(0) >0$ there can be  a spectral line shift to the blue if $\Delta_a >\Delta_\epsilon$, but when $a_s(0) <0$ the spectral line shift is always to the red.

To understand the mechanism for the formation of ultralong-range Rydberg molecules, one should consider the interaction that leads to Eq.~\ref{Fermishift}. This interaction between the Rydberg electron at position ${\bf r}$ and the perturber atom at a distance ${\bf R} = R {\bf z}$ from the Rydberg ion, is the Fermi pseudopotential, henceforth in atomic units
\begin{equation}
V_{s}({\bf r} - {\bf R}) = 2\pi  a_s(k) \delta ({\bf r} - {\bf R}).
\label{FermiPS}
\end{equation}
Here we have extended the concept of zero-energy scattering length to include a momentum dependence,  $k$. It is now possible, without resorting to solving the full molecular Schr\"odinger equation, to obtain the energy shift $E_{nlm_l}(R)$ due to $V_s({\bf r} - {\bf R})$. By employing the Rydberg atom wave function $\phi_{nlm_l}(R)$ at energy $E_{nl}$, the total energy \cite{Du1987,Du1989,greene2006},
\begin{equation}
E_{nlm_l}(R) \equiv U_{BO}(R) = E_{nl} - 2\pi a_s(k) |\phi_{nlm_l} (R)|^2.
\label{PSshift}
\end{equation}
In this equation, $n$ and $l$, are respectively, the principal and orbital angular momentum quantum numbers of the Rydberg electron at energy $E_{nlm_l}$,  while $m_l$ is the projection quantum number of $l$ onto the internuclear axis. Eq.~\ref{PSshift} includes the energy shift of the Rydberg electron due to the presence of the  perturber atom. We note here that any contributions due to higher order partial-wave interactions, and relativistic spin corrections are omitted, and discussed later in this review.

A comparison of Eq.~\ref{Fermishift} and Eq.~\ref{PSshift} offers a reinterpretation of density.  Whereas the dielectric broadening and shift of the Rydberg lines in Eq. \ref{Fermishift} depend on the density of perturbing gas particles, the energy shift in Eq.~\ref{PSshift} is proportional to the quantum mechanical electron probability density, i.e. the modulus of wave function squared at the position of the perturber atom. 

\begin{figure}
\centering
\includegraphics[width=\textwidth]{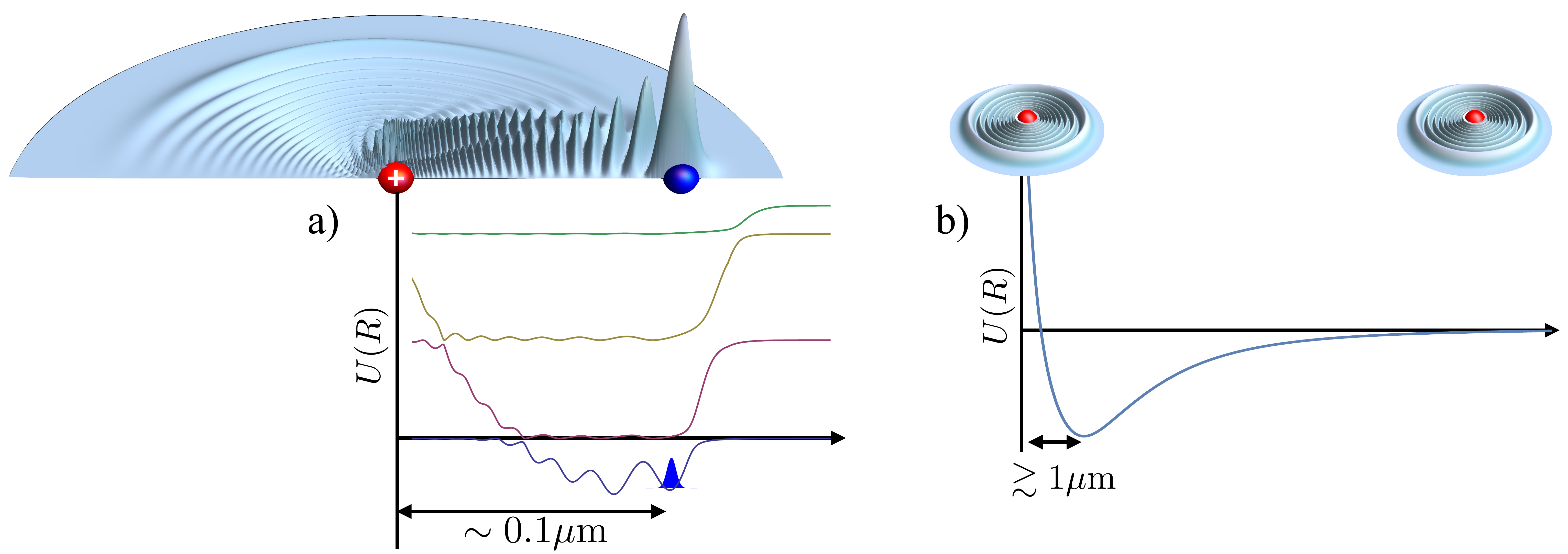}
\caption{Ultracold Rydberg molecules and macrodimers.  Schematics of Rydberg molecular electronic density and potential energy curves; (a) depicts an ultralong-range Rydberg molecule while (b) shows a Rydberg macrodimer. The ultralong-range Rydberg molecule consists of a ground state atom or a polar molecule embedded in the electronic orbital of the Rydberg atom. The potential energy curves exhibit oscillations.  A Rydberg macrodimer consists of two Rydberg atoms interacting via macroscopic electrostatic multipolar interactions. The Rydberg electron orbitals ($\sim 0.1 \text{~}\mu$m) in this schematic are not shown in proportion to the separation distances.}\label{fig:potsandwfs}  
\end{figure}

The form of Eq. \ref{PSshift} suggests that the Born-Oppenheimer (BO) potential energy curves, $U_{BO}(R)$, will be oscillatory functions of $R$, since the Rydberg wave function oscillates, see Fig.~\ref{fig:potsandwfs}. When $a_s(0)$ is negative, the potentials will possess localized wells at extremely large separations, which scale with $n^2$, and can support bound vibrational levels. Such Rydberg molecules were predicted by Greene {\it et al.} \cite{*[{}] [{. This paper was the first to propose the existence of ultra-long range Rydberg molecules.}] greene2000}.  The properties of these bound states can be predicted from the near threshold continuum scattering. The undulations in the BO potential energy curves have been confirmed with detailed quantum chemical calculations, e.g. ~\cite{Jeung1999}.  

\subsection*{Interactions between two Rydberg atoms}

When the Rydberg atom density is sufficiently high, Rydberg atoms can  interact electrostatically at large distances via  van der Waals (vdW)-type forces \cite{haroche81,gallagher98,pillet98}.  In a typical ultracold atomic gas, Rydberg-Rydberg interactions can still be substantial at internuclear separations exceeding $10 \,\mu$m \cite{singer2005,Schwettmann2006,Marcassa2014}.  The long range interactions between Rydberg atoms induce Rydberg blockade \cite{Jaksch2000}, due to the large multipolar transition moments between Rydberg states. These interactions can be expressed in an expansion in powers of $R^{-1}$,
\begin{equation}\label{multipole}
H(r_1,\theta_1;r_2,\theta_2;R) = \\ \sum_{l_1 =1}^{\infty} \sum_{l_2=1}^{\infty} \sum_{m=-l_<}^{+l_<} C_{l_1,l_2,m} \frac{r_1^{l_1} r_2^{l_2}}{R^{l_1 + l_2 + 1}} Y_{l_1}^m(\theta_1, \phi_1) Y_{l_2}^{-m}(\theta_2,\phi_2),
\end{equation}
where $C_{l_1,l_2,m}$ exist in closed form in \cite{Wolf68,Schwettmann2006,Schwettmann2007}.

In the above expression, $(r, \theta, \phi)$ are the electronic coordinates of the Rydberg electron on each atom, and $Y_{l}^{m}(\theta,\phi)$ are spherical harmonics, with $l_<$ the smaller of $l_1$ or $l_2$. Note that $l_1$ and $l_2$ are the moments of the charge distribution and not the orbital angular momenta of the electrons. For example, the set $l_1 =l_2= 1$ gives the dipole-dipole interaction. To obtain the BO potential energy curves, the Hamiltonian, consisting of the individual atomic Hamiltonians and the Rydberg-Rydberg interaction is diagonalized \cite{Schwettmann2006}. The full Hamiltonian may also include interactions with external electric and magnetic fields, which are present in experiments. The field dressed potential energy surfaces are dependent on the angle between the internuclear axis and a lab-frame quantization axis \cite{Carrol2004,Saffmann2010,Cabral2011} and $R$.  The lab-frame axis can be determined by an external field or laser polarization. 

There exist avoided crossings in the doubly-excited Rydberg BO potential curves due to the large density of states.  Each term in the  multipole expansion is $R$-dependent and can be either repulsive or attractive.  The $R$-dependence of the multipolar terms in the expansion \ref{multipole} hybridizes the atomic states as the Rydberg atoms approach, resulting in complex couplings between different molecular states. The hybridization occurs due to the electric fields produced by the Rydberg electron from one atom acting on the other. The interplay between different terms in the multipolar series can conspire to produce attractive wells at large $R$, in which Rydberg macrodimers could form, Fig.~\ref{fig:potsandwfs}. The name macrodimers is apt because of the enormous internuclear separations, $> 1 \,\mu$m \cite{Boisseau2002,Schwettmann2007,Overstreet2009,samboy2011a,samboy2011b,Sassmussen2016}. The potential wells can be controlled with small electric fields which shift the Rydberg energy levels and induce avoided crossings of the potential energy surfaces \cite{Schwettmann2007,Overstreet2009}.  

Experiments on Rydberg macrodimers are strong tests of Rydberg atom interactions and serve as a direct test for the accuracy of calculations of the doubly-excited Rydberg BO potentials. These Rydberg molecules can regulate dephasing and decay in thermal vapors for applications in electric field sensing \cite{Fan15,Kumar2016}, probe correlations in quantum gases, and generate gauge potentials \cite{Boisseau2002,kiffner2013,Marcassa2014,Buchler2015}.

\subsection*{Realization of ultralong-range Rydberg molecules }
The first realization of ultra-long range Rydberg molecules was made in an ultracold rubidium gas, in which a fraction of atoms were excited into Rb(ns) states \cite{[{}] [{. This paper describes the first observation of ultra long-range Rydberg molecules.}] bendkowsky2009}. The two-color excitation process photoassociated $^3\Sigma$ Rydberg molecules. The electronic charge density for the Rb$(5s-ns) ^3\Sigma$ Rydberg molecule should be spherical as was anticipated in Ref. \cite{greene2000}. The BO potential in this electronic state contains several vibrational bound states which were observed as peaks to the red of the atomic Rydberg Rb(35s-37s) lines. The molecular lines were measured by field ionizing the Rydberg molecules. Subsequent measurements observed ultralong-range Rydberg molecules in other species and in different initial atomic angular momentum states \cite{Tallant2012,Booth2015,Bellos2013,Anderson2014,krupp2014,Merkt2015,niederprum2016,DeSalvo2015}.

A more exotic class of homonuclear Rydberg molecules forms when excited electronic states with nearly degenerate high angular momentum components mix. This hybridization of opposite parity electronic states results from the electric field of the Rydberg electron in scattering from the perturber atom\cite{greene2000,sadeghpour2013}. The direct excitation of such molecular states with  non-zero permanent electric dipole moments (PEDM), requires a large number of photons due to the dipole selection rules. The insight that led to the observation of the trilobite molecules with the usual two-color photoassociation schemes was the realization  that some low-$l$ states in alkali-metal atoms, such as Rb($ns$) or Cs($ns$) Rydberg states, have quantum defects, $\mu_l$, with small non-integer parts \cite{rittenhouse2011}. For instance, in exciting Rb($ns$) states with quantum defect $\mu_s=3.13$, the $(n-3)l\geq 3$ manifolds with vanishing $\mu_l$ lies close in energy to the $ns$ state. An infinitesimal mixing of the degenerate $(n-3)l\geq 3$ manifold with the $ns$ state introduces an opposite parity component into the $ns$ wave function, rendering two-photon excitation of homonuclear Rydberg molecules with non-vanishing PEDM possible. 

The mixing of opposite parity degenerate hydrogenic manifold into $ns$ states in Rb was  conclusively shown to lead to formation of Rb($5s-ns \ ^3\Sigma$) Rydberg molecules with Debye-size PEDM \cite{*[{}] [{. The first observation of a permanent electric dipole moment in a homonuclear molecules due to the fractional mixture of ``trilobite'' electronic character was reported in this paper.}] li2011}. In rubidium, the mixing of the degenerate manifold into Rb($ns$) states is at best $0.1\%$. The PEDMs scale as $(n-\mu_s)^{-2}$; this scaling was confirmed in the experiments \cite{li2011}. 

\begin{figure}
\centering
\includegraphics[width=1.0\textwidth]{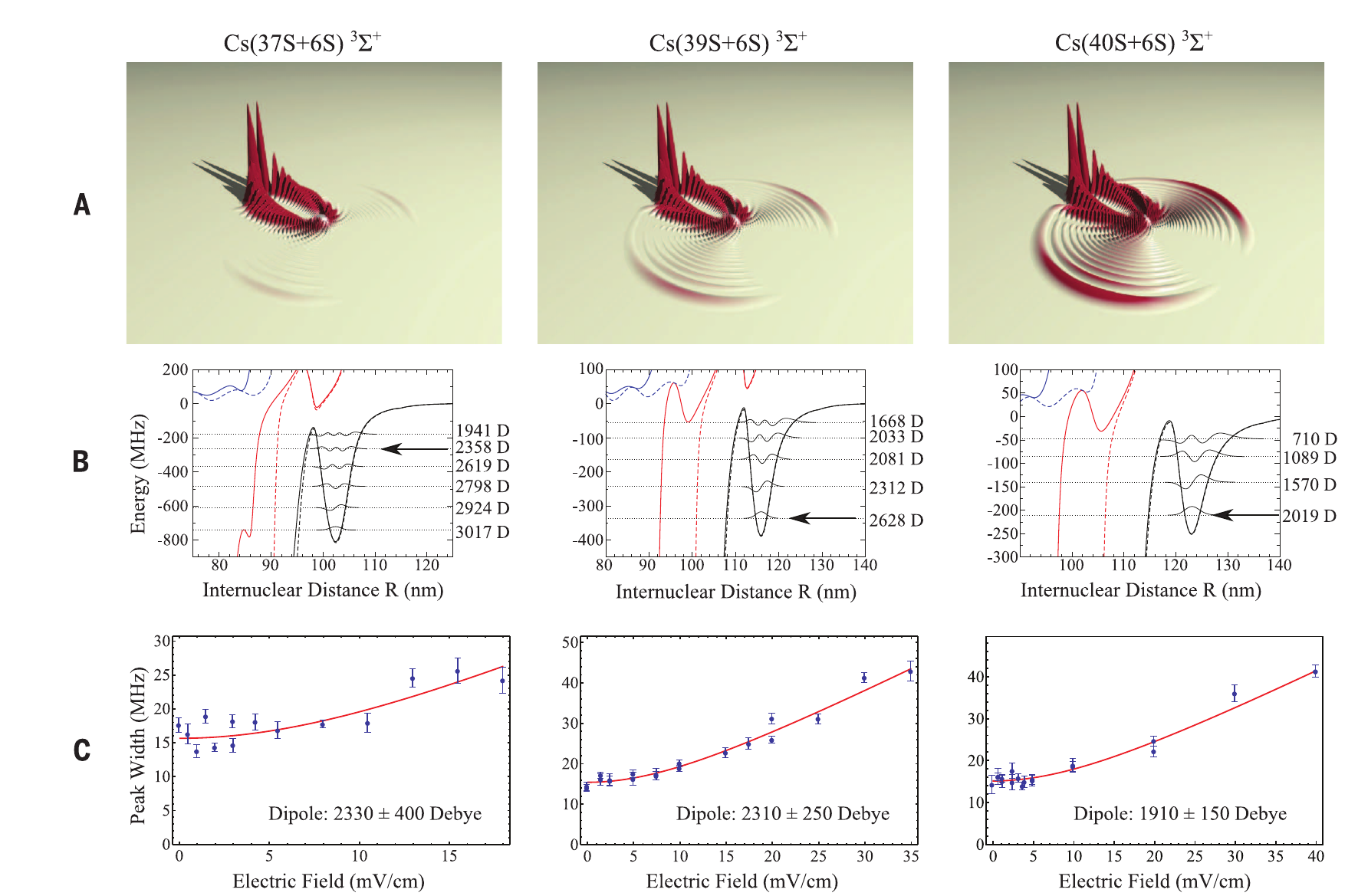}
\caption{Trilobite Rydberg molecules in an ultracold Cs gas.  a.) Calculated electron density distributions for the $^3\Sigma^+$ states of Cs correlating to $37s+6s$, $39s+6s$, and $40s+6s$. b.) The corresponding potential energy curves for the states shown in (a.) with calculated vibrational levels. c.) The broadening of the vibrational  levels indicated by an arrow in (b.) as a function of applied electric field. The changes in the lineshape as a function of electric field are used to determine the dipole moments indicated in the figure. Figure from Ref. \cite{Booth2015}.\label{fig:BoothFig}}
\end{figure}

Two unique features in Cs contribute to produce strong mixing and hence large PEDM: the Cs($ns$) small non-integer quantum defect,  $\mu_s=4.05$, causes the $(n-4) l\geq 3$ degenerate manifolds to lie energetically close to the $ns$ states, and the p-wave resonances occur at much lower energies than the corresponding resonances in Rb, mixing in more strongly the Cs($ns$) states with the close-by degenerate angular momentum manifolds, see the Section on higher-order partial wave scattering.
The resulting BO potential curves support bound vibrational levels to the blue of Cs($ns$) thresholds  with large PEDM \cite{Tallant2012}.  Observed blue-shifted spectral features were confirmed in calculations of Rydberg molecular states above the dissociation thresholds. The first trilobite Rydberg molecules were formed  in two-photon association of Cs(6s-37,39,40s $^3\Sigma$) Rydberg molecule states \cite{[{}] [{. This paper describes the first observation of trilobite molecules.}] Booth2015}, see Fig.~\ref{fig:BoothFig}. Kilo-Debye PEDM for the trilobite Rydberg molecules were measured in field ionization spectra, indicating state mixing as high as $90\%$.

\subsubsection*{Higher-order partial wave scattering}\label{highl}
The pseudopotential method in Eq.~\ref{FermiPS} has been particularly powerful in predicting the pressure broadening and shifts in high principal quantum number excitations, $n>20$, when the ground state atom involved is a noble gas.  However, chinks soon appeared in its potent armor. This situation was rectified by Omont \cite{omont1977}, some forty years after Fermi's seminal contribution \cite{fermi1934}, who showed that the origin of deviations from Fermi's prediction was in the electron scattering phase shift.  In cases with low $n$, when the electron kinetic energy, $\text{KE} = E+e^2/R$, varies as a function of the perturber position $R$, a proper description of the electron-neutral interaction must include the energy dependence of the $s$-wave scattering length, $a_s(k) = -\tan(\delta_s(k))/k$, where $k=\sqrt{2\text{ KE}}$ is the electron momentum in atomic units.

In many but not all cases, the $s$-wave contact interaction is adequate to describe the observed Rydberg molecular features \cite{bendkowsky2009,li2011}. Omont showed how to incorporate higher partial wave interactions, specifically, and most importantly, for alkali-metal atoms, where electron-scattering in the $p$-wave \text{is resonant} near threshold. In electron- alkali atom scattering, a metastable ${}^3P$ negative ion exists at fairly low energies (e.g. $\sim 0.03$ eV in rubidium \cite{Bahrim2001b,Bahrim2001a,Bahrim2000}).

The $p$-wave contribution to the electron-atom interaction potential takes the form \cite{omont1977},
\begin{equation}
V_p({\bf r})= 6 \pi a_p^3(k) \delta\left({\bf r}\right)\overleftarrow{\nabla}\cdot \overrightarrow{\nabla} \label{Pwaveint}
\end{equation} 
where $a_p^3(k)=-\tan(\delta_p(k))/k^3$ is the energy dependent $p$-wave scattering volume, and $\delta_p(k)$ is the energy dependent $p$-wave scattering phase shift. The arrows over the $\nabla$ operators indicate the direction in which the operator acts. As with the $s$-wave interaction, the strength of the $p$-wave scattering length is dependent on the kinetic energy of the electron. Note that when the scattering phase shift becomes resonant, $\delta_p(k)=\pi/2$, the $p$-wave contact interaction strength diverges.  This divergence happens exactly at the energy of a meta-stable state of an electron bound to the neutral perturber. Further details on $p$-wave scattering effects on Rydberg molecular states are discussed in {\bf Box 2}. An immediate consequence of the strong resonant $p$-wave interaction is the formation of a new class of Rydberg molecules, the so-called "butterfly" molecules \cite{hamilton2002,Chibisov2002}. The butterfly Rydberg molecules, like their larger and bigger brethren, the trilobite Rydberg molecules, possess PEDMs, albeit smaller ones. The butterfly Rydberg molecules were recently observed in Ref. \cite{niederprum2016}.

\subsubsection*{Spin coupling}
\begin{figure}
\centering
\includegraphics[width=0.8\textwidth]{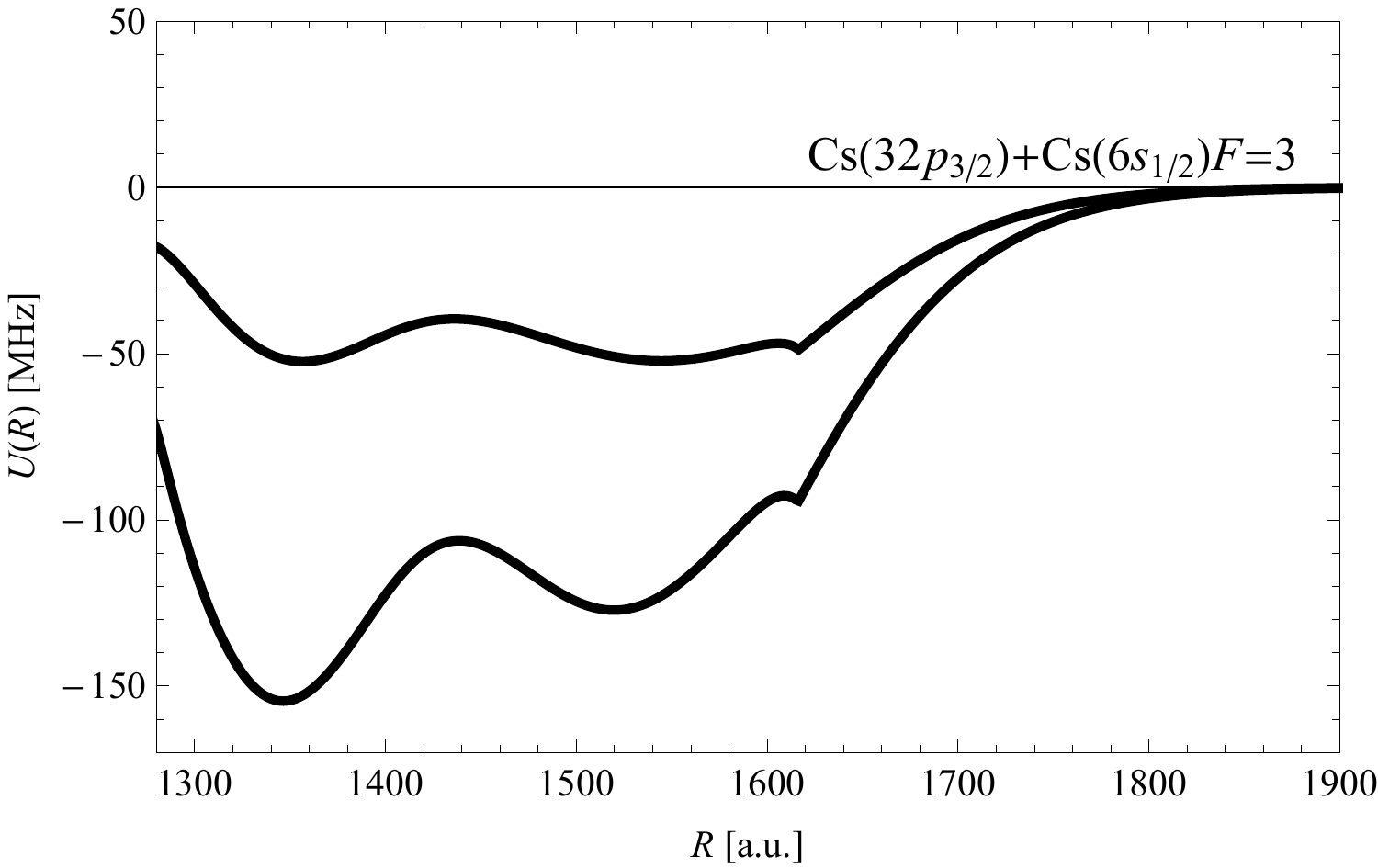}
\caption{Singlet/triplet mixing in Cs Rydberg molecule BO potential energy curves. The interaction potentials including spin-dependent relativistic interactions (see {\bf Box 3} for details). The lowest potential is mainly a $S=1$ (triplet) BO potential for a Cs($6s_{1/2}$) - Cs($32p_{3/2}$) Rydberg molecule in the ground hyperfine state, F=3. The excited BO potential curve is predominantly a spin-mixed $S=0$ and $S=1$ (singlet and triplet) state. Unphysical cusps can be seen in both potentials at $R\sim 1629$ a.u.  The cusps are the result of the semi-classical description of the energy-dependent scattering length at the transition from the classically allowed to the classically forbidden region.  The cusps {cannot be removed at the level of approximation discussed in this review, but they } have little influence on the nature of molecular states of interest here. These potentials were recently employed in Ref. \cite{Markson2016} for interpretation of the observation of spin-mixed states in Ref. \cite{Merkt2015}.}
\label{fig:spinmxing}
\end{figure}

In the preceding discussion, interactions between a ground state atom and a Rydberg electron were assumed to be spin-independent. While the scattering phase shifts did separately depend on the spin of the electrons, the scattering amplitudes added up incoherently.  In alkali metal atoms, the two valence electrons couple into singlet and triplet spinors: ${\bf{S}=\bf{s}_r+\bf{s}_g}= 0, {\rm{and}} \, 1$, with $\bf{s}_r$ and $\bf{s}_g$, respectively, the Rydberg atom electron and ground atom electron spins. The spin-orbit (SO) interaction couples the Rydberg orbital angular momentum $\bf{l}_r$ to $\bf{s}_r$, with a coefficient $A_{\text{SO}}$.  The hyperfine (HF) interaction couples the nuclear spin of the ground state atom, $\bf{i}_g$, to $\bf{s_g}$, with an isotropic coefficient $A_{\text{HF}}$. The HF interaction in the Rydberg states and SO interaction in the Rydberg degenerate manifolds are ignored \cite{haroche81}. The scattering electron angular momentum, $\bf{L}$, also couples to $\bf{S}$ and gives rise to partial-wave scattering amplitudes  with the total electron spinor, ${\bf J}= {\bf L} + {\bf S}$. For $p$-wave scattering, i. e. $L = 1$, three fine-structure phase shifts ($J=0, \, 1, \, 2$) result. Typical BO interaction potential energies which result from including the relativistic spin terms are shown in Fig.~\ref{fig:spinmxing}, where a new interaction curve whose spin character is mixed singlet/triplet, emerges.

The details of the spin-dependent Hamiltonian and the different interaction terms are given in {\bf Box 3}. The first calculations of Rydberg molecules, including the hyperfine interaction, but neglecting singlet scattering, were performed by Anderson {\it et al.} \cite{*[{}] [{. This paper was the first to predict the existence of spin mixed potentials in diatomic Rydberg molecules.}] Anderson2014} for Rb. Already, in the observed spectrum of Rb(35,37d$_{5/2}$)F=2 Rydberg molecules \cite{Raithel2014}, there were features which matched the $v=0$ vibrational levels in the mixed singlet/triplet potential energy curves. The observation and interpretation of singlet/triplet mixing due to the Cs ground-state HF interaction was made in Ref.~\cite{Merkt2015}. This interpretation neglected the important $p$-wave scattering contribution in Cs, which was remedied in Ref.~\cite{Markson2016}. Mixed molecular states were observed and interpreted in Rb as well \cite{*[{}] [{. This paper reported the first observation of p-wave interaction dominated ``butterfly'' molecules.}] niederprum2016,niederprum2016a,Kleinbach2017,Bottcher2016,Eiles2017}.  Kleinbach {\it et al.} \cite{Kleinbach2017} observed blue-shifted states in Rb due to resonant spin-orbit effects; Tallant {\it et al.} \cite{Tallant2012} had observed the blue-shifted states in Cs due to strong non-adiabatic effects.

\subsection*{Realization of Rydberg macrodimers}
Rydberg macrodimers,  Fig.~\ref{fig:potsandwfs}, are formed by pairing Rydberg atoms with attractive long range and repulsive shorter range interactions such that potential wells can form. Due to the large polarizability of Rydberg atoms and high density of states, the bond lengths can be larger than $1\,\mu$m. The potential wells that bind the atoms together are broad on the molecular scale with well depths ranging from MHz to GHz. The multipolar interactions between pairs of Rydberg atoms that lead to the formation of macrodimers  were treated in detail in \cite{singer2005,Schwettmann2006,Schwettmann2007}. Broader description of Rydberg atom-Rydberg atom interactions can be found in \cite{Marcassa2014,Cabral2011,Pillet2010,Saffmann2010,Gallagher2008,Jones2012}.

\begin{figure}
\centering
\includegraphics[width=\textwidth]{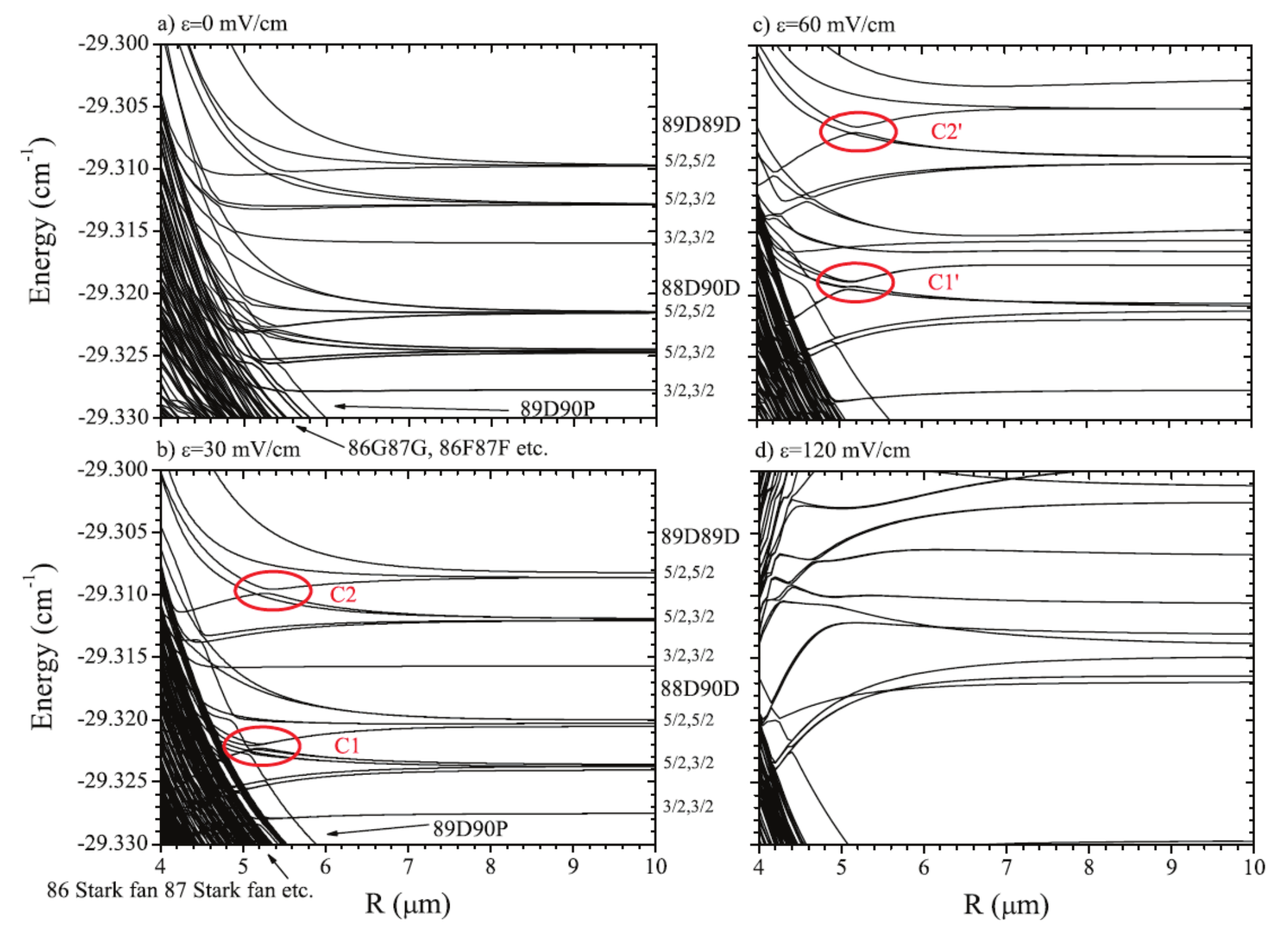}
\caption{Cs Rydberg macrodimer BO potential energy curves. The Cs Rydberg macrodimer potentials in the region of the $89d+89d$ asymptote in the presence of an electric field, $\varepsilon$. The total electronic angular momentum projection quantum number is m=3. The electric field lies along the internuclear axis. Some potential energy curve crossings that support bound states are labeled, $C1$, $C2$, $C1^\prime$, and $C2^\prime$. The labels along the vertical axes give the angular momentum $j$ for each atom of the pair. Figure from  \cite{Schwettmann2007}. }\label{fig6}
\end{figure}

Rydberg macrodimers were first predicted by Boisseau {\it et al.} \cite{*[{}] [{. This paper describes the first prediction of the existence of macrodimers.}] Boisseau2002}. Subsequent calculations were carried out in \cite{Schwettmann2007,samboy2011a,samboy2011b,kiffner2012,kiffner2013}.  Macrodimers were first observed by Overstreet {\it et al.} \cite{[{}] [{. This paper describes the first experimental observation of macrodimers.}]Overstreet2009} and later by Sa{\ss}mannshausen {\it et al.} \cite{Sassmussen2016}.  Rydberg macrotrimers have also been predicted \cite{samboy2013}.

In the original Rydberg macrodimer description \cite{Boisseau2002}, interatomic potential energy wells supporting bound states were calculated to exist for a combination of attractive quadrupole ($R^{-5}$) and repulsive vdW ($R^{-6}$ and $R^{-8}$) potentials, Fig.~\ref{fig6}.  The equilibrium distance for these molecules scales as $0.3 n^3 a_0$; for $n\sim 70$ the equilibrium distances are greater than $1\,\mu$m.  The size of the Rydberg electronic wave function is smaller than the equilibrium distance, i.e. the potential wells lie outside the LeRoy radius \cite{Boisseau2002,Schwettmann2006}. As a consequence, exchange interactions between the Rydberg electrons can be neglected. In the case of Rydberg macrodimers the LeRoy radius is overly restrictive because the orbitals of the electrons are so diffuse. 

A full description of Rydberg macrodimer formation requires that interactions between different potential energy curves be incorporated. The Rydberg macrodimers form as attractive and repulsive potentials undergo avoided crossings \cite{Schwettmann2007}. The avoided crossings can occur naturally among potential curves with the same symmetry, or with an external field, Fig.~\ref{fig6}. Complicated state crossings, arising from the interplay of different multipolar terms in the Hamiltonian, allow for tuning of the resonant Rydberg macrodimer potentials with small electric fields. Background fields of $< 1\,$V$\,$cm$^{-1}$ are necessary for Rydberg states around $n> 50$ \cite{Cabral2011,Nascimento2009}. Because the potential wells can be induced at larger $R$, the well depths are shallow, scaling as $n^{-3}$, while the equilibrium distances scale as $n^{8/3}$.  As an illustrative example, this places the equilibrium distance $> 5\,\mu$m for $n\approx90$, Fig.~\ref{fig6} \cite{samboy2011a,samboy2011b}.

\begin{figure}
\centering
\includegraphics[width=\textwidth]{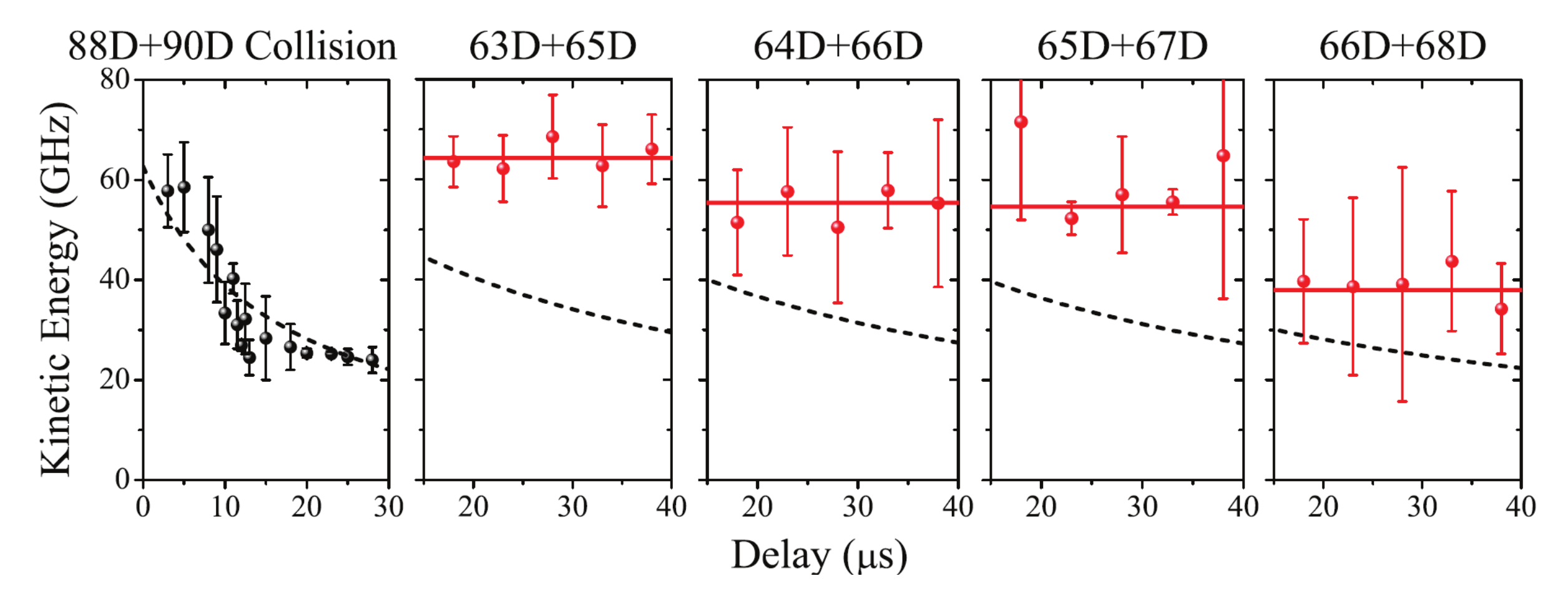}
\caption{Kinetic energy spectra from the fragmentation of macrodimers. The kinetic energy as a function of the delay, $\tau$,  between when a particular molecular complex was excited and when it was probed via field ionization. In the leftmost panel a state correlating to $88d+90d$ is first excited. The other figures correspond to states correlating with $63d+65d$, $64d+66d$, $65d+67d$ and $66d+68d$. The dashed line in the leftmost panel shows the kinetic energy expected for a pair of Cs atoms photoassociated at $R=2.8\mu$m and moving uniformly apart at a velocity of $17\,$cm$\,$s$^{-1}$. The other four panels show the same quantity for macrodimer states for an electric field of $224\,$mV$\,$cm$^{-1}$ for $63d+65d$,   $205\,$mV$\,$cm$^{-1}$ for $64d+66d$, $190\,$mV$\,$cm$^{-1}$  for $65d+67d$ and $158\,$mV$\,$cm$^{-1}$ for $66d+68d$.  The red points are the data while the dashed line is what is expected for a laser induced collision of the pair of Cs atoms with a recoil energy equal to the thermal energy of the atoms in the ultracold atom trap. Figure from \cite{Overstreet2009}.}
\label{fig7}
\end{figure}

The first observation of Rydberg macrodimers was by photoassociation and subsequent recoil ion momentum spectroscopy in ultracold Cs($n \sim 60$) \cite{Tallant2006,Overstreet2007,Overstreet2009}. Cs($n\sim 41$) Rydberg macrodimer states were later observed spectroscopically \cite{Sassmussen2016}.  One of the difficulties in observing macrodimers is that the molecular potential wells are shallow and broad when compared to typical diatomic molecules. The vibrational frequencies are small, $< 10\,$MHz at $n\sim 40$ and $< 100\,$kHz at $n\sim 90$.  As a consequence,  the spectra of a Rydberg macrodimer appears as a continuum when spectral broadening effects such as rotation, radiative and blackbody decay, and laser spectral bandwidths are taken into account. Normally, rotation of a macrodimer can be neglected since the rotational constant of such a large molecule is small, $\sim 3\,$Hz at $n\sim 90$, but because a large number of rotational states can be populated, there will be rotational broadening which causes the vibrational states to overlap at larger $n$ \cite{Schwettmann2007}.

A continuous spectrum is not necessarily a difficulty, but in the case of macrodimers it becomes a challenge because the Rydberg atom interaction potentials are so flat that it becomes possible to excite Rydberg pairs on repulsive curves, inducing a photodissociative collision. The observed features in the spectra of ultracold Rydberg gases that asymptotically corresponded to the excitation of Rydberg atom pairs, or states shifted away from atomic Rydberg states, have been difficult to interpret, in particular to identify macrodimers\cite{Browaeys2013,farooqi2003,Overstreet2007,singer2005,stanojevic2008,kiffner2012,Yu2013}.  Recoil ion momentum spectroscopy, Fig.~\ref{fig7} is able to distinguish the different cases, since photoinduced collisions convert potential energy of the complex into kinetic energy of the product atoms, while bound states remain at relatively well-defined $R$ over the lifetime of the molecule. The signatures of these different behaviors can be distinguished by measuring the ion recoil after field ionization of the Rydberg atoms \cite{Overstreet2007,Overstreet2009,Marcassa2014}, Fig.~\ref{fig7}. The spectral lineshapes of Cs Rydberg macrodimers have now been observed for $n\sim 41$ \cite{Sassmussen2016}. This results in a lineshape where a peak exists at the potential minimum, decreases with vibrational excitation and increases as the Rydberg density of states grows at the dissociation limit \cite{Schwettmann2007,Sassmussen2016}.

\subsection*{Few-body and many-body Rydberg molecule spectroscopy}
\begin{figure}
\centering
\includegraphics[width=2.5in]{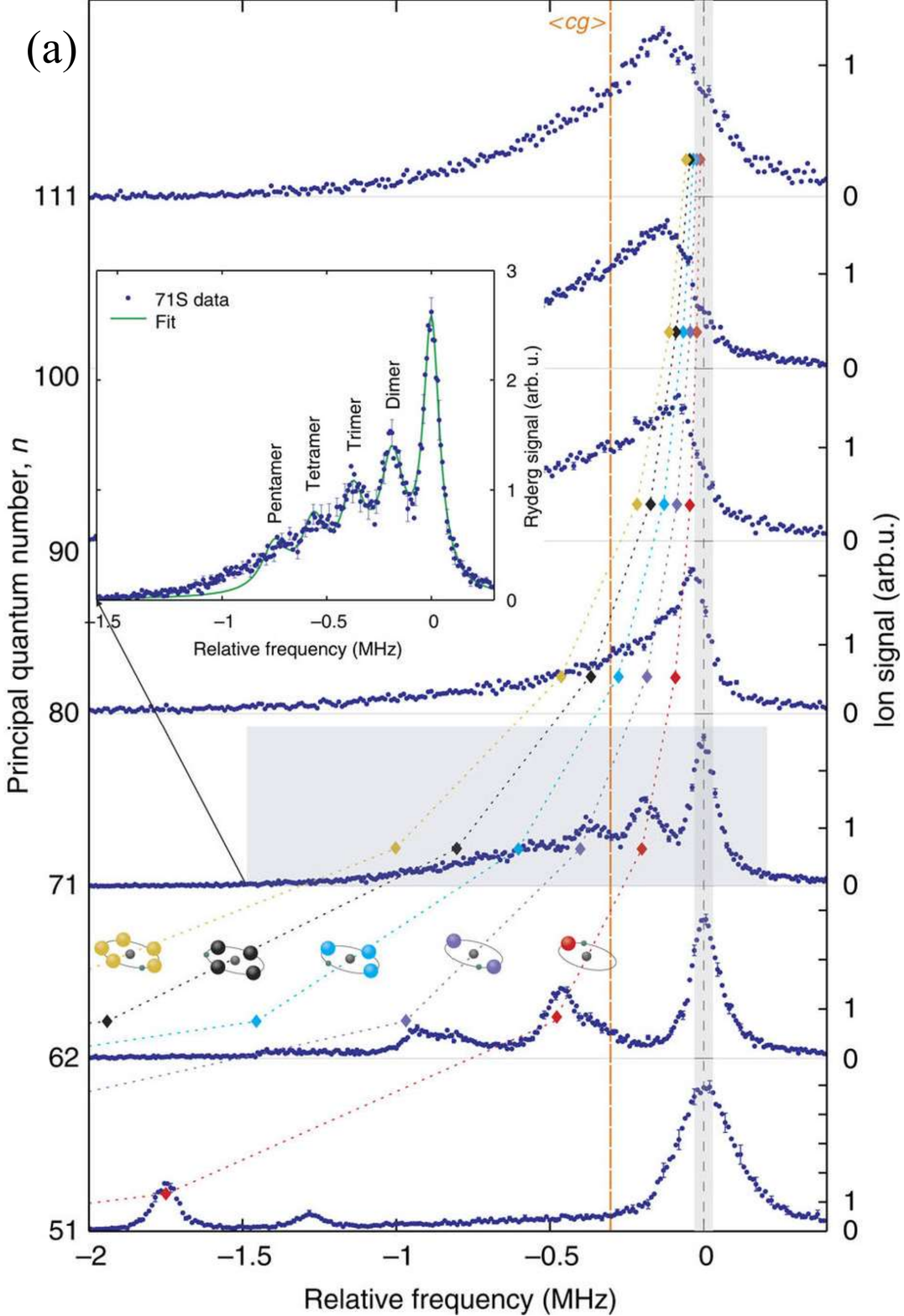}
\includegraphics[width=3.2in]{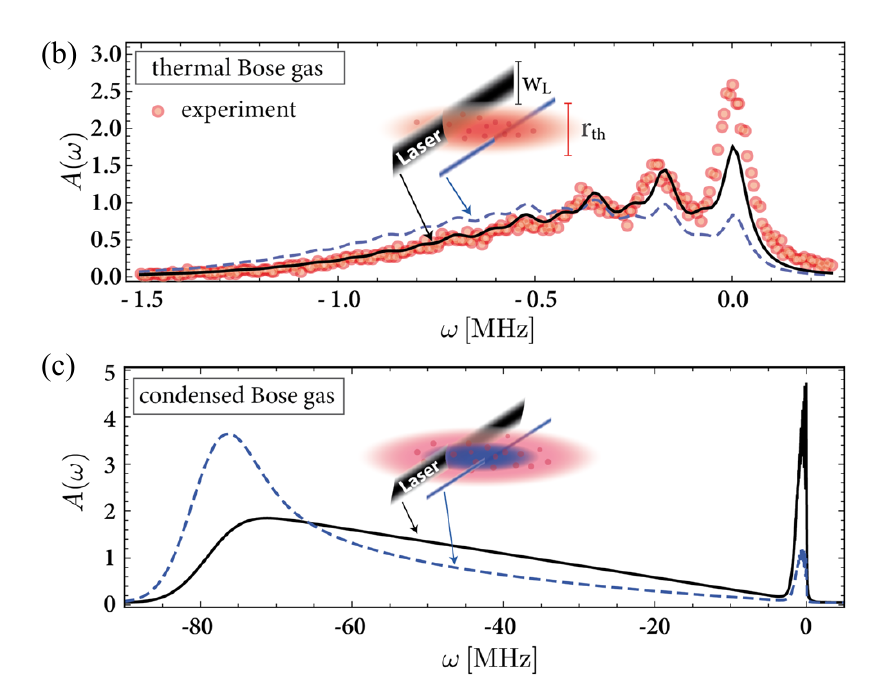}
\caption{Rb Rydberg dimer, trimer, tetramer, and many-body spectra. (a) The observation of polyatomic $s$-wave Rydberg molecules (dimers, trimers, and tetramers) in photoassociation of Rydberg molecules; with permission from \cite{gaj2014}, (b) density averaged absorption spectra for Rb(71s) excitations. In (a), the gas is thermal at T =0.5 $\mu$K and the MOT center density is $\rho =1.7 \times10^{12}$ cm$^{−3}$. The laser profile illuminates the cloud over the full cylindrical radius (black) and in a small waist (dashed, blue); data is reported in Gaj {\it et al.} \cite{gaj2014}. In (b), the density averaged absorption spectrum at T/T$_c = 0.47$ for a partially condensed gas of (zero temperature) at peak density $\rho = 2.3 × 10^{14}$ cm$^{−3}$. The figure is adopted with permission from \cite{Schmidt2016}. }
\label{fig:polaron}
\end{figure}

Even at the atomic densities used in the first observation of  ultralong-range Rydberg molecules \cite{bendkowsky2009,Bendkowsky2010}, there were hints that bound molecular states with more than one ground state atom could form. Calculations \cite{Liu2006,liu2009} showed that an additional ground state atom in the Rydberg electron orbit would bind to make a triatomic molecule in homonuclear systems such as with Rb or Cs. With increasing gas density, or increasing $n$, larger oligomeric molecules (tetramers, pentamers, ...) may form \cite{Eiles2016,Rios2016}. These higher order molecular states of Rb$(5s-ns ^3\Sigma)$  were observed in \cite{*[{}] [{. The observation of larger oligomeric bound states with multiple ground state atoms bound to a single Rydberg atom was first reported in this paper.}] gaj2014,Liebisch2016}, see Fig.~\ref{fig:polaron}a. Stretching and bending vibrations and electric field effects on triatomic Rydberg molecules were also studied \cite{Fey2016,Fernandez2016}. Gaj {\it et al.} \cite{gaj2014} found that with higher $n$ Rydberg excitations, $n \sim 80-110$, so many perturber atoms could exist in a Rydberg orbit that a spectral overlap of molecular lines produced a density dependent broadening, $\Delta = \frac{2\pi \hbar^2 a_s}{m}\int{|\Psi(R)|^2}\rho(R)dR = \frac{2\pi \hbar^2 a_s}{m}{\bar \rho}$, where $\rho(R)$ is the local density from the position of the Rydberg ion and $\bar \rho$ is the weighted average density, Fig.~\ref{fig:polaron}a.

A theory for the excitation of giant impurities in a quantum gas was developed by Schmidt {\it et al.} \cite{Schmidt2016} using the functional determinant approach (FDA), Fig.~\ref{fig:polaron}b-c. This time-dependent method is capable of unmasking multiscale time dynamics in a quantum gas by accounting for quasiparticle excitations and  molecular formation. The FDA simulations revealed the resolved molecular oligomeric excitations in a Rb BEC \cite{schlagmuller2016}, as well as the unresolved broad features at much larger detuning, heralding the crossover from few-body to many-body polaronic states. 

In Rydberg excitation in an ultracold gas, large electron scattering phase shifts due to the presence of triplet $p$-wave shape resonances \cite{Bahrim2000,hamilton2002} introduce complicated non-adiabatic interactions between BO potential energy curves. A classical Monte Carlo simulation, sampling the energy shift from each Rydberg-perturber distance in the BO potential energies, can describe the observed spectral profiles \cite{schlagmuller2016}. However, this classical interpretation, which relies on $p$-wave resonances to produce large detuning shifts in the interaction potentials, cannot adequately describe the large $n$ spectral tails emanating from the highest density condensed regions. The narrowing of the spectral tails is described numerically and analytically in terms of formation of a class of bosonic polarons, the Rydberg polarons. The classical interpretation misses the underlying  many-particle Rydberg molecule formation.

In heavy alkaline-earth atoms (calcium, strontium and barium), the electron scattering in the $p$-wave channel binds (positive electron affinity) and there are no resonances. This atomic system provides a clean platform to study many-body Rydberg impurity dressing. Ultralong-range Sr Rydberg dimers have been observed \cite{DeSalvo2015}.  The lifetimes of these Sr Rydberg molecules are longer than comparable states in Rb due to the aforementioned absence of $p$-wave shape resonances \cite{Camargo2016}.

An FDA prediction for the formation of Rydberg polarons in a uniform density gas is the appearance of a Gaussian density profile, Fig.~\ref{fig:polaron}b-c \cite{Schmidt2016}. It was shown that the Gaussian width narrows with increasing $n$, or the local density \cite{Schmidt2017,Camargo2017}, as $\Gamma \propto \sqrt{\rho_0}/n^3$, with $\rho_0$ the peak uniform density in the BEC. The emergence of the Gaussian profile is understood by decomposing the spectral line function in terms of excitations from a ground state BEC to interacting single-particle Rydberg molecule states. A binomial decomposition, including only two states, yields a Gaussian profile in the limit of large particle number. The observation of Rydberg excitations in Sr in \cite{Camargo2017} and the detailed theory of FDA in \cite{Schmidt2017} reveal the intrinsic spectrum of Rydberg polarons which cannot be explained by mean field or classical interpretations \cite{Schmidt2017}.

The examples given in this section, show that the extreme properties of Rydberg atoms and their controllability can be used for experiments on many- and few- body physics. This line of investigation has only recently emerged.

\subsection*{Rydberg dimer lifetimes}
Ultralong-range Rydberg molecular lifetimes for Rb, Cs and Sr have been measured \cite{bendkowsky2009,Butscher2011,Booth2015,Camargo2017}. In Rb Rydberg molecules, two observed inelastic channels are a) Rb$_2^+$ molecular ions, and b) $l$-mixing reactions \cite{SChlagmuller2016b,Sassmannshausen2016}. Both reactions are largely due to the presence of strong $p$-wave shape resonances, producing an acceleration of the internuclear motion to the short-range "chemical" region (see {\bf Box 2} for details). The Rydberg core - atom interaction and the butterfly potential avoided crossings at short distances with degenerate manifold trilobite states produce the molecular ion and $l$-mixed states. At high principal quantum numbers, $90 < n < 110$, a sudden increase in the lifetime of the Rydberg molecules was observed \cite{SChlagmuller2016b}. This threshold increase in lifetime could be due to many-body interaction of the Rydberg atom with the surrounding Bose gas and not influenced by the butterfly or trilobite interactions.

The absence of $p$-wave resonances in electron scattering phase shifts from Sr($5s$) atoms translates to generally longer lifetimes for Sr-photoassociated Rydberg molecules \cite{Camargo2017}.  Even in higher density gases, where the average occupation number of perturber atoms in a Rydberg orbit is large,  the molecular lifetimes are longer, on the order of the the Rydberg atom lifetime \cite{Whalen2017}.

In Cs, for intermediate $n$ quantum number excitations, the strong mixing of Cs($ns$) states with the degenerate Cs($(n-4)l\geq 3$) manifolds, which produces trilobite molecules with kilo-Debye PEDM, dictates that the lifetimes are controlled by the highest nearly quantum-defect free angular momentum states. The Cs($5s-ns ^3\Sigma$) molecular lifetimes were found to be influenced by the Cs($(n-4)f$) state lifetimes \cite{Booth2015}. 

The lifetimes of the Rydberg macrodimers are limited by the atomic Rydberg lifetimes, with spontaneous and blackbody decay  \cite{Schwettmann2007,samboy2011a,Marcassa2014}. Non-adiabatic processes and ionization can also limit the molecular lifetime \cite{Sassmannshausen2016}, but in many cases the contribution of these decay mechanisms is much smaller than the atomic radiative lifetime \cite{Schwettmann2007}.  Electronic autoionization is small because the equilibrium separation of the Rydberg atoms is typically large enough than there is little or no overlap of the electronic wavefunctions with the respective Rydberg ions. A quick look at Fig.~\ref{fig:potsandwfs} illustrates the relative size of the macrodimer equilibrium separation in comparison to the extent of the electronic wave function. Additionally, the spread of the Rydberg electron wave function and the short-range repulsive walls limit ionization. As long as the inner turning point in the bound state potential is outside the region where potentials correlating to high angular momentum Rydberg states are, as shown in Fig.~\ref{fig6},  the lifetimes of macrodimers will be on the order of the Rydberg atomic lifetimes. Calculations for Cs(89d) Rydberg macrodimers gave a liftetime of $\sim450\,\mu$s, determined entirely by Rydberg atom decay \cite{Schwettmann2007}. Measurements at densities of $\sim 10^{10}\,$cm$^{-3}$ for photoinduced collisions and for macrodimers have affirmed this picture \cite{Overstreet2007,Overstreet2009}. In cases where the the inner well can strongly interact with the curves with high angular momentum character, Penning ionization can occur at a high rate, leading to molecular decay \cite{Merkt2015b}.  In ultracold gases where the density is larger, $\sim 10^{12}\,$cm$^{-1}$, shorter macrodimer lifetimes have been observed \cite{Sassmussen2016}. The shorter lifetime was attributed to the presence of a third, nearby atom.

\subsection*{Giant polyatomic Rydberg molecules}
The description in the Section on collisional interactions between a Rydberg atom and a ground-state atom involving the interaction of a Rydberg electron and a ground-state perturber atom invoked the concept of zero-range scattering, i. e. $a_s(0)$. When the perturber is a dipolar molecule- a diatomic molecule with a PEDM, such as OH or KRb- the scattering length picture no longer holds. The first-order of interaction is charge-dipole  \cite{*[{}] [{. The existence of yet be observed Rydberg molecules consisting of an atom in a highly excited Rydberg state and a ground-state polar molecule were first predicted here.}] rittenhouse2010}, 
\begin{equation}
V_{ed}({\bf r} - {\bf R}) = B{\bf N}^2 - {\bf d} \cdot {\bf F ({\bf r - {\bf R})}},
\label{eq:e-dipole}
\end{equation}
where ${\bf d}$ is the dipole moment of the molecular perturber, and ${\bf N}$ and $B$ are the rotational angular momentum and constant for a rigid rotor molecule, e. g. KRb. The electric field due to the Rydberg electron is ${\bf F} =\frac{e ({\bf r} - {\bf R})}{|{\bf r} - {\bf R}|^3}$, which polarizes the molecular dipole. For dipole moments above the critical dipole, known as the Fermi-Teller dipole moment, $d > d_{\mathrm{cr}} \approx 1.63$ D, the electron may bind to the molecule, forming a negatively charged ion. When $d < d_{\mathrm{cr}}$, the anisotropic interaction of the Rydberg electron with the PEDM can form polyatomic Rydberg molecular states. These types of BO potential energy curves exhibit long-range wells and oscillatory behavior.  Molecular states as described here have been calculated for a range of molecules with PEDM, e. g. $\Lambda$ doublet and rigid-rotor molecules. These polyatomic potentials display two outer wells, whose electronic wave functions correspond to orientation of the molecular dipole toward and away from the Rydberg core \cite{rittenhouse2010}.

Ultracold giant polyatomic Rydberg molecules have been studied in the context of electric field control \cite{mayle2012}, coherent control of orientation \cite{rittenhouse2010,rittenhouse2011}, hybridization of rotational states of dipolar molecules \cite{Gonzalez2015}, and for conditional addressing and non-destructive readout of rotational quantum bits \cite{Kuznetsova2011,Kuznetsova2016}. Polyatomic Rydberg molecules can potentially be realized using atomic Rb (or K) Rydberg excitation in an ultracold gas of KRb molecules. 

\section*{Summary and outlook}
Macrodimers and ultralong-range Rydberg molecules have exotic features: macroscopic bond length, kilo-Debye PEDM, tunability with laboratory sized fields, and fascinating electronic probability densities (trilobites and butterflies). Due to large equilibrium sizes, their rotational constants are in Hz-kHz regimes, and in experiments, localized rotational wave packets are excited. Efforts on both observational and theoretical fronts have evolved to utilize such distinct molecular properties for probes of quantum gases, for spin alignment and control, and for selective chemical reactivity in the ultracold regime. Cases where the Rydberg atom core is used to trap an electron inside a quantum gas are one example \cite{Balewski2013}.  Spin-mixing of molecular potentials in ultralong-range Rydberg molecules allows for control of spin interactions and may open new vistas on the atomic analogue of the celebrated central spin problem, in which a central spin interacts with a large number of strongly coupled spins in the environment. Rydberg molecules have already been investigated as probe of atomic occupation in optical lattices for quantum gas microscopy \cite{Ott2015}. The many-body macroscopic occupation of Rydberg molecules in a BEC has been realized \cite{Schmidt2016,Camargo2017}, but probing Pauli blocking in the many-body spectral Rydberg profile is worthy of investigation. Recently, it's been proposed that trilobite Rydberg molecules still persist in random dense ultracold atomic gases \cite{Luukko2017}.

Formation of Rydberg molecules places fundamental limits on applications in sensing and quantum information with Rydberg atoms. The interaction of Rydberg electrons with gas atoms can decohere electromagnetically induced transparency Rydberg polaritons and limit Rydberg blockade, light storage and photon-photon interactions in ultracold Rydberg gases \cite{Maxwell2013,Thompson2017,Mirgorodskiy2017}.  Rydberg macrodimer formation and Rydberg atom collisions can be a major contributor to dephasing and loss in experiments that require coherent light-matter interaction \cite{Fan15,Kumar2016}. 

A major reason for investigating ultracold Rydberg molecules is their novel properties, such as their large dipole moments and relatively long lifetimes. Ultracold Rydberg molecules have binding energies which are much less than their internal electronic energies. Because these two degrees of freedom are weakly coupled, ultracold Rydberg molecules can live for relatively long times allowing them to be used as spectroscopic probes of quantum gases.  These properties might also be leveraged to use ultra-long range Rydberg molecules as an intermediate state for formation of optical Feshbach resonance molecules \cite{Sandor2017,Thomas2017}. The extreme nature of ultralong-range molecules can give insights into quantum chemistry, as was studied in coherent photodissoaction and photoassociation with rotary echo techniques \cite{Butscher2010}.  Butterfly molecules \cite{niederprum2016}, by virtue or their favorable Franck-Condon overlap with low-lying Rydberg states, can be utilized to populate vibrationally bound heavy Rydberg systems, i. e. ion-pairs of ultracold atoms \cite{Kirrander2013,Markson2016a}. 

The Rydberg electrons are nearly perfect traps for the ions, so hybrid ion-atom trapping interactions could be envisioned with Rydberg molecules. It may be possible, with suitable density of Rydberg excitations, to realize ultralong-range Rydberg molecules and Rydberg macrodimers in the same trap, perhaps engineering the best properties of each type of molecule for tailored investigations. Another possible line of inquiry is to use Rydberg macrodimers to investigate gauge potentials \cite{Wilczek1986,Cederbaum1989,Baer2002}. Rydberg macrodimers have relatively easily controlled interaction potentials, so they offer interesting possibilities to engineer these interactions. Imaging particle dynamics can determine the exit channel trajectories if the initial state is aligned in the lab frame, thus measuring the effects of a pseudo-magnetic field (pseudo-Lorentz force on the colliding particles).

Ultracold Rydberg molecules are vehicles for applying low-energy quantum scattering techniques to the spectroscopy of quantum gases. Rydberg atom-Rydberg atom interactions are already receiving increased attention as a way to engineer collective states for many applications. As new kinds of exotic molecules, the exaggerated properties of Rydberg molecules are sure to be harnessed for new experiments.   
\newpage
\begin{tcolorbox}[breakable, enhanced,title=Box 1]
The interaction between a Rydberg electron and a ground state atom is, in the first order perturbation theory, a polarization potential $
W(q_i) = -\sum_i{\frac{e^2\alpha}{2q_i^4}}$, 
where $\alpha$ is the polarizability of the ground state particles, and $q_i$ is the electron position from the i$^{\rm th}$ perturber core.  All spectral lines are hence expected to be redshifted from the unperturbed levels, as the polarization potential for a perturber in a non-degenerate ground state is always attractive. The shift of the Rydberg lines due to dielectric screening is obtained in cm$^{-1}$ as in Eq. 23 in \cite{fermi1934}, 
\begin{equation}
\Delta_\epsilon \approx 2.6{(\frac{4\pi}{3})}^{4/3} [\frac{(\epsilon -1)}{16\pi^2\hbar  c}] e^2 \rho^{1/3},
\label{polshift}
\end{equation}
where $\epsilon$ is the dielectric constant, normalized to vacuum permittivity and $\rho$ is the gas density.  
Fermi determined that this simple picture was flawed because the Rydberg electron-perturber interaction is fundamentally a scattering process.
\end{tcolorbox}
\newpage

\begin{tcolorbox}[breakable, enhanced,title=Box 2]

The effect of $p$-wave electron-atom scattering on the molecular Born-Oppenheimer potentials of an ultralong-range Rb$_2$ Rydberg molecule is shown in  Fig.~\ref{fig:RbpotsPwave}.   A set of (on this scale) flat potentials exist at each Rydberg energy threshold. Plunging down through these potentials, a $p$-wave dominated potential is seen starting at the $n=30$ hydrogenic threshold. This potential cuts through the 33s, 31d, and 32p thresholds before turning back up at shorter range.  This new potential corresponds to an attractive $1/R$ Coulomb potential between the Rydberg Rb$^+$ core and a metastable Rb$^-$(6s6p) ion.  

The fact that the $p$-wave scattering is not isotropic, leads to an interaction that depends on the 3-dimensional gradient of the Rydberg molecule wave function at the location of the perturber atom.  The $p$-wave interaction maximizes this gradient of the electron wave function.  The fact that the derivative acts in all three spatial dimensions gives rise to two types of potentials; those where the derivative is maximum along the internuclear axis resulting in $\Sigma$ type molecules whose potentials are illustrated in Fig. \ref{fig:RbpotsPwave}(a), and those where the derivative is maximum perpendicular to this axis resulting in $\Pi$ type molecules whose potentials are illustrated in Fig.~\ref{fig:RbpotsPwave}(b).   $\Sigma$ molecules also experience the $s$-wave electron-perturber interaction meaning that the deep $p$-wave potentials also couple to $s$-wave dominated potentials seen near each Rydberg threshold.  $\Pi$ type molecules have a zero in the radial portion of the Rydberg electron wavefunction, and thus do not experience the s-wave electron-perturber interaction.

As the ionic potential plunges down, it goes through a broad crossing with the next nearly degenerate manifold of high angular momentum ($l>2$) hydrogenic Rydberg states ($n=29$ in the case of Fig.~\ref{fig:RbpotsPwave}).  The upper branch of this crossing is highlighted in the inset of Fig.~\ref{fig:RbpotsPwave}(a).  
In the $\Sigma$ Rydberg molecules, the radial Rydberg oscillation forms wells that are deep  enough to bind localized vibrational states creating a new kind of $p$-wave interaction dominated molecule, the so called ``butterfly'' molecules \cite{hamilton2002,Chibisov2002}.

Fabrikant {\it et al.} \cite{khuskivadze2002}, included the different fine structure terms of the $^3P_J$ phase shifts in the $ns ^3\Sigma$ states. The contributions from the $J=0$ and $J=2$ phase shifts in the $ns ^3\Sigma$ curves produce oscillations in the potential curves, with mainly $\Sigma$ molecular character, while the potential energy curve resulting from the $^3P_1$ scattering retains mainly a $\Pi$ molecular state, diabatically crossing other curves. The mostly $\Sigma$ curves in the intermediate-$R$ regions support the butterfly Rydberg molecular states.

\begin{center}
\includegraphics[width=3in]{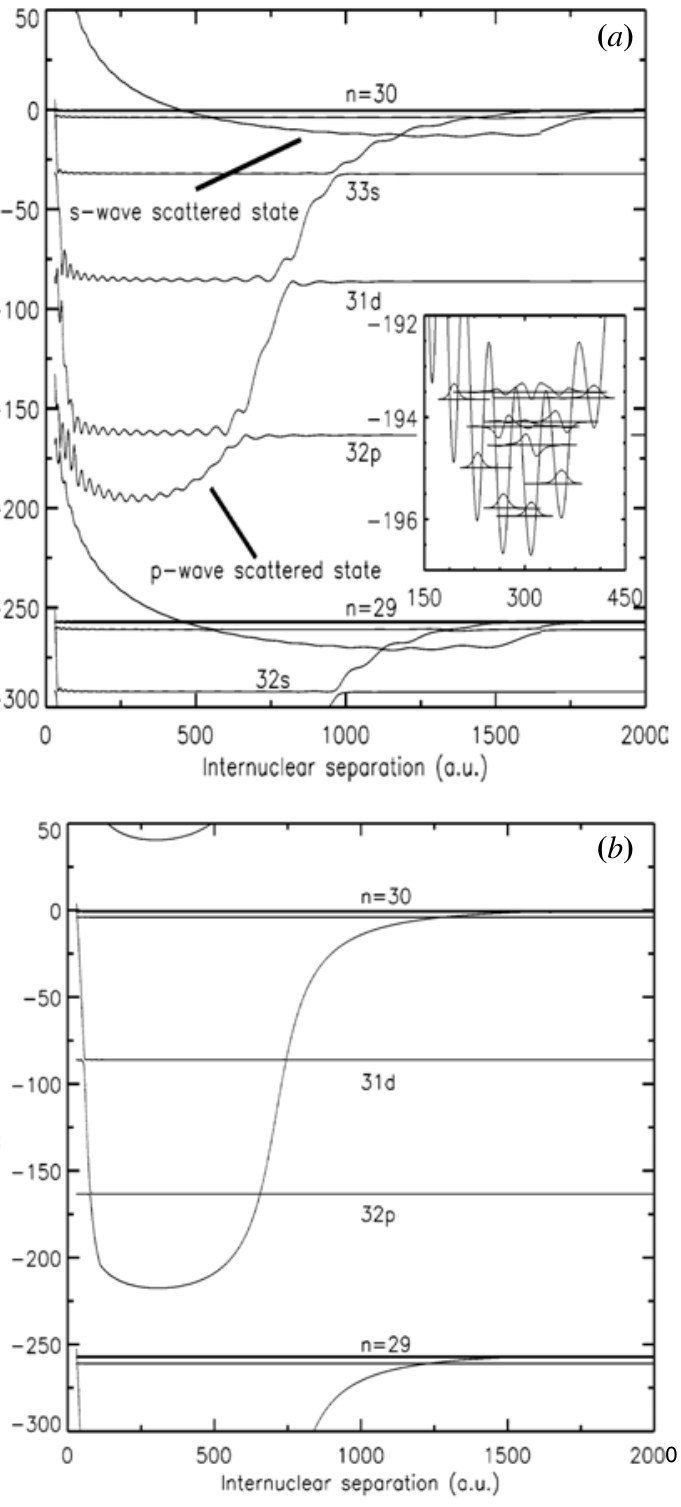}
\end{center}
\captionof{figure}{P-wave electron-atom interactions. The Rb Rydberg dimer potentials near the Rb(31d) threshold are shown for cases where the $p$-wave interaction is along (a), and perpendicular to (b) the internuclear axis.  In (a) the normal Rydberg molecule potentials dominated by the $s$-wave electron-Rubidium interaction go through a set of narrow avoided crossings with a $p$-wave interaction dominated potential that falls sharply as the internuclear separation decreases.  Similar behavior can be seen in (b), however, because a node has been enforced in the angular portion of the Rydberg electron interaction, there is no $s$-wave electron-atom interaction and hence no oscillation in the potentials. The inset in (a) shows a new set of $p$-wave dominated potentials and the vibrational states created by them which correspond to the butterfly type Rydberg molecule. Figure from Ref. \cite{hamilton2002}.}\label{fig:RbpotsPwave}

\end{tcolorbox}

\newpage

\begin{tcolorbox}[breakable, enhanced,title=Box 3]


The relevant spin-dependent relativistic terms in the Hamiltonian for the ultracold Rydberg-atom ground-atom system have been incorporated in successive approximations \cite{khuskivadze2002,Raithel2014,Markson2016,Eiles2017}.  All of these contributions are encapsulated in the Hamiltonian,
\begin{align}
\hat{\bf H} = & \hat{\bf H}_{0}+ A_{\text{SO}}{\bf l}_{r}\cdot{\bf s}_{r}%
+A_{\text{HF}}{\bf i}_{g}\cdot{\bf s}_{g}+\sum_{S,M_{S}}2\pi a_{s}^{\left(
S\right)  }\left(  k\right)  \delta\left(  \bf{r}-\bf{R}\right)  \left\vert
SM_{S}\right\rangle \left\langle SM_{S}\right\vert \label{Eq:angcoupling} \\
& +\sum_{J,S,M_J,M_{L},M_{L}^{\prime}} 6\pi\left[  a_{p}^{\left(
S,J\right)  }\left(  k\right)  \right]  ^{3}C_{LM_{L};S\left(  M_J
-M_{L}\right)  }^{JM_J}C_{LM_{L}^{\prime};S\left(  M_J
-M_{L}^{\prime}\right)  }^{JM_J}\delta\left(  \bf{r}-\bf{R}\right) \nonumber\\
& \qquad \qquad \times \left\vert S,M_J-M_{L}\right\rangle \overleftarrow{\nabla}^{\left(
M_{L}\right)  }\cdot\overrightarrow{\nabla}^{\left(  M_{L}^{\prime}\right)
}\left\langle S,M_J-M_{L}^{\prime}\right\vert \nonumber.
\end{align}
from Ref. \cite{Eiles2017}. Here, $M_L$ is the projection of the electron scattering angular momentum on the internuclear axis; for Rb and Cs Rydberg excitations, scattering of the Rydberg electron from the ground state Rb or Cs atoms produces $M_L=0 \, (\Sigma)$ and $M_L=1 \, (\Pi)$ molecular symmetries, and $\overleftrightarrow{\nabla}^{(M_L)}$ is the $M_L=0,\pm 1$ spherical component of the gradient operator acting in the indicated direction.  The total spin angular momentum projection is given by $M_S$.  The projection of spatial and spin electron angular momentum along the inter-nuclear axis is given by $M_J = M_L + M_S$, 

$\hat{\bf H}_0$ is the bare Rydberg Hamiltonian and the other terms in $\hat{\bf{H}}$ describe pair-wise couplings between the various angular momentum quantities present in the system,  shown schematically in Fig. \ref{fig:SpinFig}.  The first coupling term is the spin-orbit interaction for the Rydberg electron and the next coupling term is the hyperfine interaction in the ground state atom.  The remaining terms are sums over different internal angular momentum states, and incorporate the $s$- and $p$-wave electron-perturber interactions which are dependent on the total electronic spin ${\bf S} = {\bf s}_r+{\bf s}_g$. 

Because there is a non-zero relative orbital angular momentum between the Rydberg electron and the ground-state perturber, there is an additional spin-orbit effect.  This additional spin-orbit effect results in a ${\bf J = L+S}$ dependent $p$-wave interaction that couples the total spin with the relative orbital angular momentum ${\bf L}$ \cite{khuskivadze2002,Markson2016,Eiles2017}.   It should be noted here that within the $J$ dependent, $p$-wave interaction there is a subtle additional coupling between relative orbital angular momentum ${\bf L}$ and the total electron spin, ${\bf S}$, that results from these two vectors precessing about ${\bf J}$ during the interaction.  This coupling was first noted in Ref. \cite{khuskivadze2002} and later reformulated in the form of contact interactions as discussed in Ref. \cite{Eiles2017}.  The precession leads to additional coupling terms between the $\Sigma$ and $\Pi$ type molecular potentials that becomes strong near the $p$-wave interaction resonances.  One result of these pairwise angular-momentum coupling terms is that the only conserved angular momentum (in the body-fixed molecular frame) is the total projection along the internuclear axis, $\Omega=M_J+m_{i_g}$, where $m_{i_g}$ is the magnetic projection of the ground state nuclear spin.

\begin{center}
\includegraphics[width=0.8\textwidth]{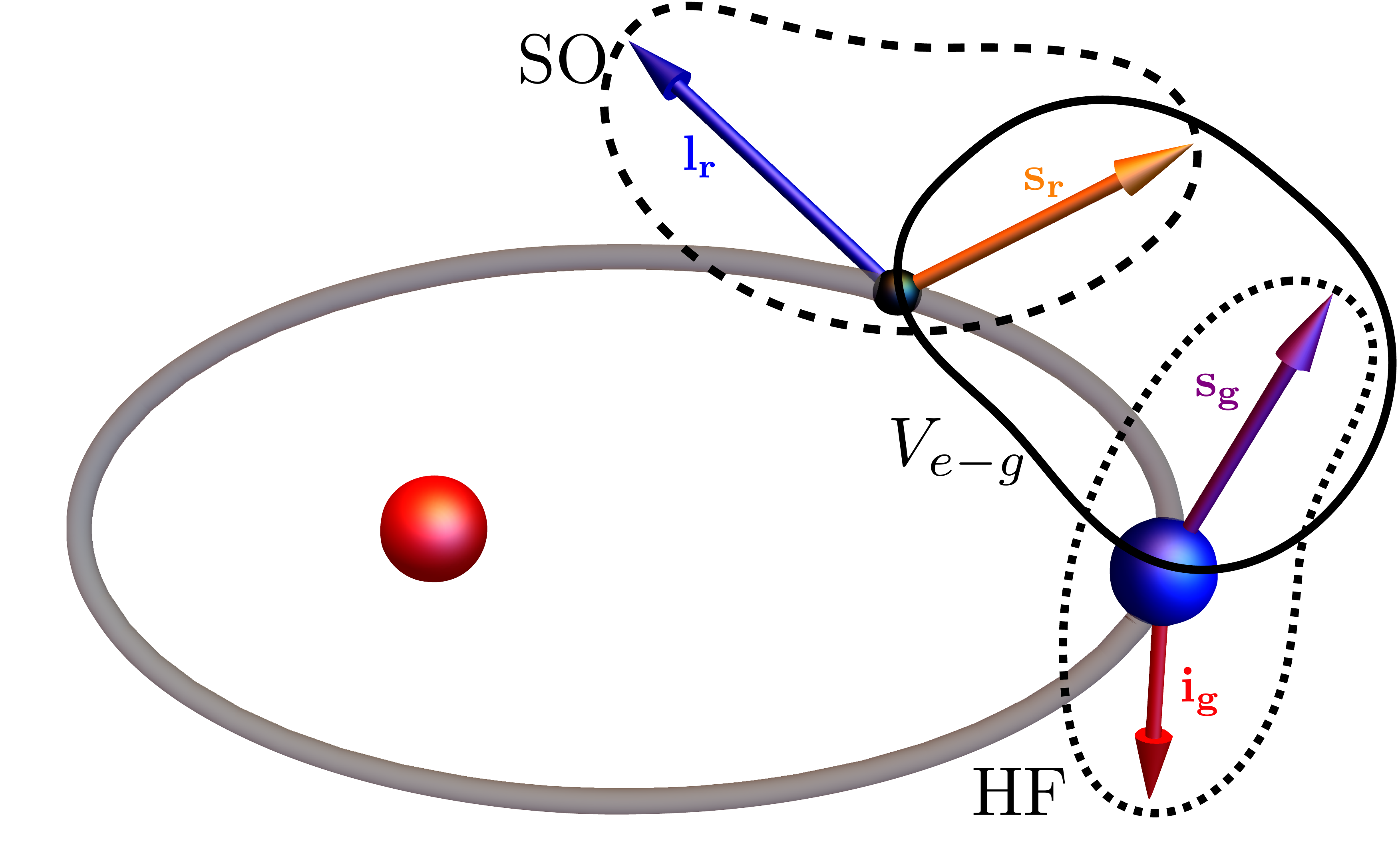}
\end{center}
\captionof{figure}{Pairwise angular momentum coupling. A diagram of relevant spin coupling terms in Rydberg atom - ground atom interaction. The red and blue balls indicate the positions of the Rydberg ionic core and the ground perturber atom, respectively. $V_{e-g}$ signifies the spin-dependent scattering interactions. All other spin terms are indicated. Because the Rydberg molecules form at large internuclear distances, hyperfine interaction in the ground perturber atom can be thought of inducing long-range spin-mixing in the molecule. Here, the dashed curve encloses the spins coupled by the spin-orbit interaction, the dotted curve encloses the spins coupled by the hyperfine interaction, and the solid curve encloses the spins correlated by the spin-dependent Rydberg electron-ground state atom interaction.}\label{fig:SpinFig}

\end{tcolorbox}

\newpage

\section*{Competing Interests}
The authors have no competing financial interests that influence this work.

\section*{Author Contributions}
All the authors contributed equally to this work.

\section*{Acknowledgements}
Observatory. S.T.R. acknowledges support from NSF Grant No. PHY-1516421. J.S. acknowledges support from NSF Grant No. PHY-1607296 and AFOSR grant FA9550-15-1-0381. H. R. S. acknowledges support through an NSF grant to ITAMP at Harvard University and the Smithsonian Astrophysical Observatory.

\bibliographystyle{naturemag}
\bibliography{naturecomrev}

\begin{thebibliography}{100}
\expandafter\ifx\csname url\endcsname\relax
  \def\url#1{\texttt{#1}}\fi
\expandafter\ifx\csname urlprefix\endcsname\relax\def\urlprefix{URL }\fi
\providecommand{\bibinfo}[2]{#2}
\providecommand{\eprint}[2][]{\url{#2}}

\bibitem{Balmer1885}
\bibinfo{author}{Balmer, J.~J.}
\newblock \bibinfo{title}{Notiz \"uber die spektrallinien des wasserstoffs}.
\newblock \emph{\bibinfo{journal}{Annalen der Physik und Chemie}}
  \bibinfo{pages}{80--87} (\bibinfo{year}{1885}).

\bibitem{Rydberg1889}
\bibinfo{author}{Rydberg, J.~H.}
\newblock \emph{\bibinfo{journal}{Den Kungliga Svenska Vetenskapsakademiens
  Handlingar}} \textbf{\bibinfo{volume}{23}}, \bibinfo{pages}{11}
  (\bibinfo{year}{1889}).

\bibitem{gallagher2005rydberg}
\bibinfo{author}{Gallagher, T.~F.}
\newblock \emph{\bibinfo{title}{{R}ydberg atoms}}
  (\bibinfo{publisher}{Cambridge University Press}, \bibinfo{year}{2005}).

\bibitem{amaldi1934}
\bibinfo{author}{Amaldi, E.} \& \bibinfo{author}{Segre, E.}
\newblock \bibinfo{title}{Effetto della pressione sui termini elevati degli
  alcalini}.
\newblock \emph{\bibinfo{journal}{Il Nuovo Cimento (1924-1942)}}
  \textbf{\bibinfo{volume}{11}}, \bibinfo{pages}{145--156}
  (\bibinfo{year}{1934}).

\bibitem{fermi1934}
\bibinfo{author}{Fermi, E.}
\newblock \bibinfo{title}{Sopra lo spostamento per pressione delle righe
  elevate delle serie spettrali}.
\newblock \emph{\bibinfo{journal}{Il Nuovo Cimento (1924-1942)}}
  \textbf{\bibinfo{volume}{11}}, \bibinfo{pages}{157--166}
  (\bibinfo{year}{1934}).

\bibitem{Du1987}
\bibinfo{author}{Du, N.~Y.} \& \bibinfo{author}{Greene, C.~H.}
\newblock \bibinfo{title}{Interaction between a {R}ydberg atom and neutral
  perturbers}.
\newblock \emph{\bibinfo{journal}{Phys. Rev. A}} \textbf{\bibinfo{volume}{36}},
  \bibinfo{pages}{971--974} (\bibinfo{year}{1987}).
\newblock \urlprefix\url{https://link.aps.org/doi/10.1103/PhysRevA.36.971}.

\bibitem{Du1989}
\bibinfo{author}{Du, N.~Y.} \& \bibinfo{author}{Greene, C.~H.}
\newblock \bibinfo{title}{Multichannel {R}ydberg spectra of the rare gas
  dimers}.
\newblock \emph{\bibinfo{journal}{The Journal of Chemical Physics}}
  \textbf{\bibinfo{volume}{90}}, \bibinfo{pages}{6347--6360}
  (\bibinfo{year}{1989}).
\newblock \urlprefix\url{http://dx.doi.org/10.1063/1.456352}.

\bibitem{greene2006}
\bibinfo{author}{Greene, C.~H.}, \bibinfo{author}{Hamilton, E.~L.},
  \bibinfo{author}{Crowell, H.}, \bibinfo{author}{Vadla, C.} \&
  \bibinfo{author}{Niemax, K.}
\newblock \bibinfo{title}{Experimental verification of minima in excited
  long-range rydberg states of ${\mathrm{rb}}_{2}$}.
\newblock \emph{\bibinfo{journal}{Phys. Rev. Lett.}}
  \textbf{\bibinfo{volume}{97}}, \bibinfo{pages}{233002}
  (\bibinfo{year}{2006}).
\newblock
  \urlprefix\url{https://link.aps.org/doi/10.1103/PhysRevLett.97.233002}.

\bibitem{greene2000}
\bibinfo{author}{Greene, C.~H.}, \bibinfo{author}{Dickinson, A.~S.} \&
  \bibinfo{author}{Sadeghpour, H.~R.}
\newblock \bibinfo{title}{Creation of polar and nonpolar ultra-long-range
  {R}ydberg molecules}.
\newblock \emph{\bibinfo{journal}{Phys. Rev. Lett.}}
  \textbf{\bibinfo{volume}{85}}, \bibinfo{pages}{2458--2461}
  (\bibinfo{year}{2000}).
\newblock \urlprefix\url{https://link.aps.org/doi/10.1103/PhysRevLett.85.2458}.
\newblock \bibinfo{note}{\\{\bf This paper was the first to predict the
  existence of ultralong-range Rydberg molecules.}}

\bibitem{Jeung1999}
\bibinfo{author}{Yiannopoulou, A.}, \bibinfo{author}{Jeung, G.-H.},
  \bibinfo{author}{Park, S.~J.}, \bibinfo{author}{Lee, H.~S.} \&
  \bibinfo{author}{Lee, Y.~S.}
\newblock \bibinfo{title}{Undulations of the potential-energy curves for highly
  excited electronic states in diatomic molecules related to the atomic orbital
  undulations}.
\newblock \emph{\bibinfo{journal}{Phys. Rev. A}} \textbf{\bibinfo{volume}{59}},
  \bibinfo{pages}{1178--1186} (\bibinfo{year}{1999}).
\newblock \urlprefix\url{https://link.aps.org/doi/10.1103/PhysRevA.59.1178}.

\bibitem{haroche81}
\bibinfo{author}{Raimond, J.~M.}, \bibinfo{author}{Vitrant, G.} \&
  \bibinfo{author}{Haroche, S.}
\newblock \bibinfo{title}{Spectral line broadening due to the interaction
  between very excited atoms: 'the dense {R}ydberg gas'}.
\newblock \emph{\bibinfo{journal}{Journal of Physics B: Atomic and Molecular
  Physics}} \textbf{\bibinfo{volume}{14}}, \bibinfo{pages}{L655}
  (\bibinfo{year}{1981}).
\newblock \urlprefix\url{http://stacks.iop.org/0022-3700/14/i=21/a=003}.

\bibitem{gallagher98}
\bibinfo{author}{Anderson, W.~R.}, \bibinfo{author}{Veale, J.~R.} \&
  \bibinfo{author}{Gallagher, T.~F.}
\newblock \bibinfo{title}{Resonant dipole-dipole energy transfer in a nearly
  frozen {R}ydberg gas}.
\newblock \emph{\bibinfo{journal}{Phys. Rev. Lett.}}
  \textbf{\bibinfo{volume}{80}}, \bibinfo{pages}{249--252}
  (\bibinfo{year}{1998}).
\newblock \urlprefix\url{https://link.aps.org/doi/10.1103/PhysRevLett.80.249}.

\bibitem{pillet98}
\bibinfo{author}{Mourachko, I.} \emph{et~al.}
\newblock \bibinfo{title}{Many-body effects in a frozen {R}ydberg gas}.
\newblock \emph{\bibinfo{journal}{Phys. Rev. Lett.}}
  \textbf{\bibinfo{volume}{80}}, \bibinfo{pages}{253--256}
  (\bibinfo{year}{1998}).
\newblock \urlprefix\url{https://link.aps.org/doi/10.1103/PhysRevLett.80.253}.

\bibitem{singer2005}
\bibinfo{author}{Singer, K.}, \bibinfo{author}{Stanojevic, J.},
  \bibinfo{author}{Weidem{\"u}ller, M.} \& \bibinfo{author}{C{\^o}t{\'e}, R.}
\newblock \bibinfo{title}{{Long-range interactions between alkali {R}ydberg
  atom pairs correlated to the ns--ns, np--np and nd--nd asymptotes}}.
\newblock \emph{\bibinfo{journal}{Journal of Physics B: Atomic, Molecular and
  Optical Physics}} \textbf{\bibinfo{volume}{38}}, \bibinfo{pages}{S295}
  (\bibinfo{year}{2005}).
\newblock
  \urlprefix\url{http://iopscience.iop.org/article/10.1088/0953-4075/38/2/021}.

\bibitem{Schwettmann2006}
\bibinfo{author}{Schwettmann, A.}, \bibinfo{author}{Crawford, J.},
  \bibinfo{author}{Overstreet, K.~R.} \& \bibinfo{author}{Shaffer, J.~P.}
\newblock \bibinfo{title}{{Cold Cs {R}ydberg-gas interactions}}.
\newblock \emph{\bibinfo{journal}{Phys. Rev. A}} \textbf{\bibinfo{volume}{74}},
  \bibinfo{pages}{020701} (\bibinfo{year}{2006}).
\newblock \urlprefix\url{https://link.aps.org/doi/10.1103/PhysRevA.74.020701}.

\bibitem{Marcassa2014}
\bibinfo{author}{Marcassa, L.~G.} \& \bibinfo{author}{Shaffer, J.~P.}
\newblock \bibinfo{title}{Chapter two - interactions in ultracold {R}ydberg
  gases} \textbf{\bibinfo{volume}{63}}, \bibinfo{pages}{47 -- 133}
  (\bibinfo{year}{2014}).
\newblock
  \urlprefix\url{http://www.sciencedirect.com/science/article/pii/B978012800129500002X}.

\bibitem{Jaksch2000}
\bibinfo{author}{Jaksch, D.} \emph{et~al.}
\newblock \bibinfo{title}{Fast quantum gates for neutral atoms}.
\newblock \emph{\bibinfo{journal}{Phys. Rev. Lett.}}
  \textbf{\bibinfo{volume}{85}}, \bibinfo{pages}{2208--2211}
  (\bibinfo{year}{2000}).
\newblock \urlprefix\url{https://link.aps.org/doi/10.1103/PhysRevLett.85.2208}.

\bibitem{Wolf68}
\bibinfo{author}{Wolf, W.~P.} \& \bibinfo{author}{Birgeneau, R.~J.}
\newblock \bibinfo{title}{Electric multipole interactions between rare-earth
  ions}.
\newblock \emph{\bibinfo{journal}{Phys. Rev.}} \textbf{\bibinfo{volume}{166}},
  \bibinfo{pages}{376--382} (\bibinfo{year}{1968}).
\newblock \urlprefix\url{https://link.aps.org/doi/10.1103/PhysRev.166.376}.

\bibitem{Schwettmann2007}
\bibinfo{author}{Schwettmann, A.}, \bibinfo{author}{Overstreet, K.~R.},
  \bibinfo{author}{Tallant, J.} \& \bibinfo{author}{Shaffer, J.~P.}
\newblock \bibinfo{title}{{Analysis of long-range Cs {R}ydberg potential
  wells}}.
\newblock \emph{\bibinfo{journal}{Journal of Modern Optics}}
  \textbf{\bibinfo{volume}{54}}, \bibinfo{pages}{2551--2562}
  (\bibinfo{year}{2007}).
\newblock \urlprefix\url{http://dx.doi.org/10.1080/09500340701584076}.

\bibitem{Carrol2004}
\bibinfo{author}{Carroll, T.~J.}, \bibinfo{author}{Claringbould, K.},
  \bibinfo{author}{Goodsell, A.}, \bibinfo{author}{Lim, M.~J.} \&
  \bibinfo{author}{Noel, M.~W.}
\newblock \bibinfo{title}{Angular dependence of the dipole-dipole interaction
  in a nearly one-dimensional sample of {R}ydberg atoms}.
\newblock \emph{\bibinfo{journal}{Phys. Rev. Lett.}}
  \textbf{\bibinfo{volume}{93}}, \bibinfo{pages}{153001}
  (\bibinfo{year}{2004}).
\newblock
  \urlprefix\url{http://link.aps.org/doi/10.1103/PhysRevLett.93.153001}.

\bibitem{Saffmann2010}
\bibinfo{author}{Saffman, M.}, \bibinfo{author}{Walker, T.~G.} \&
  \bibinfo{author}{M\o{}lmer, K.}
\newblock \bibinfo{title}{Quantum information with {R}ydberg atoms}.
\newblock \emph{\bibinfo{journal}{Rev. Mod. Phys.}}
  \textbf{\bibinfo{volume}{82}}, \bibinfo{pages}{2313--2363}
  (\bibinfo{year}{2010}).
\newblock \urlprefix\url{https://link.aps.org/doi/10.1103/RevModPhys.82.2313}.

\bibitem{Cabral2011}
\bibinfo{author}{Cabral, J.~S.} \emph{et~al.}
\newblock \bibinfo{title}{{Effects of electric fields on ultracold {R}ydberg
  atom interactions}}.
\newblock \emph{\bibinfo{journal}{Journal of Physics B: Atomic, Molecular and
  Optical Physics}} \textbf{\bibinfo{volume}{44}}, \bibinfo{pages}{184007}
  (\bibinfo{year}{2011}).
\newblock \urlprefix\url{http://stacks.iop.org/0953-4075/44/i=18/a=184007}.

\bibitem{Boisseau2002}
\bibinfo{author}{Boisseau, C.}, \bibinfo{author}{Simbotin, I.} \&
  \bibinfo{author}{C\^ot\'e, R.}
\newblock \bibinfo{title}{Macrodimers: Ultralong range {R}ydberg molecules}.
\newblock \emph{\bibinfo{journal}{Phys. Rev. Lett.}}
  \textbf{\bibinfo{volume}{88}}, \bibinfo{pages}{133004}
  (\bibinfo{year}{2002}).
\newblock
  \urlprefix\url{https://link.aps.org/doi/10.1103/PhysRevLett.88.133004}.
\newblock \bibinfo{note}{\\{\bf This paper describes the first prediction of
  the existence of macrodimers.}}

\bibitem{Overstreet2009}
\bibinfo{author}{Overstreet, K.~R.}, \bibinfo{author}{Schwettmann, A.},
  \bibinfo{author}{Tallant, J.}, \bibinfo{author}{Booth, D.} \&
  \bibinfo{author}{Shaffer, J.~P.}
\newblock \bibinfo{title}{Observation of electric-field-induced cs {R}ydberg
  atom macrodimers}.
\newblock \emph{\bibinfo{journal}{Nature Physics}}
  \textbf{\bibinfo{volume}{5}}, \bibinfo{pages}{581 -- 585}
  (\bibinfo{year}{2009}).
\newblock \urlprefix\url{http://dx.doi.org/10.1038/nphys1307}.
\newblock \bibinfo{note}{\\{\bf This paper describes the first experimental
  observation of macrodimers.}}

\bibitem{samboy2011a}
\bibinfo{author}{Samboy, N.}, \bibinfo{author}{Stanojevic, J.} \&
  \bibinfo{author}{C\^ot\'e, R.}
\newblock \bibinfo{title}{{Formation and properties of {R}ydberg macrodimers}}.
\newblock \emph{\bibinfo{journal}{Phys. Rev. A}} \textbf{\bibinfo{volume}{83}},
  \bibinfo{pages}{050501} (\bibinfo{year}{2011}).
\newblock \urlprefix\url{https://link.aps.org/doi/10.1103/PhysRevA.83.050501}.

\bibitem{samboy2011b}
\bibinfo{author}{Samboy, N.} \& \bibinfo{author}{C{\^o}t{\'e}, R.}
\newblock \bibinfo{title}{Rubidium {R}ydberg macrodimers}.
\newblock \emph{\bibinfo{journal}{Journal of Physics B: Atomic, Molecular and
  Optical Physics}} \textbf{\bibinfo{volume}{44}}, \bibinfo{pages}{184006}
  (\bibinfo{year}{2011}).
\newblock
  \urlprefix\url{http://iopscience.iop.org/article/10.1088/0953-4075/44/18/184006}.

\bibitem{Sassmussen2016}
\bibinfo{author}{Sa\ss{}mannshausen, H.} \& \bibinfo{author}{Deiglmayr, J.}
\newblock \bibinfo{title}{Observation of {R}ydberg-atom macrodimers:
  Micrometer-sized diatomic molecules}.
\newblock \emph{\bibinfo{journal}{Phys. Rev. Lett.}}
  \textbf{\bibinfo{volume}{117}}, \bibinfo{pages}{083401}
  (\bibinfo{year}{2016}).
\newblock
  \urlprefix\url{https://link.aps.org/doi/10.1103/PhysRevLett.117.083401}.

\bibitem{Fan15}
\bibinfo{author}{Fan, H.} \emph{et~al.}
\newblock \bibinfo{title}{Atom based rf electric field sensing}.
\newblock \emph{\bibinfo{journal}{Journal of Physics B: Atomic, Molecular and
  Optical Physics}} \textbf{\bibinfo{volume}{48}}, \bibinfo{pages}{202001}
  (\bibinfo{year}{2015}).
\newblock \urlprefix\url{http://stacks.iop.org/0953-4075/48/i=20/a=202001}.

\bibitem{Kumar2016}
\bibinfo{author}{Kumar, S.}, \bibinfo{author}{Sheng, J.},
  \bibinfo{author}{Sedlacek, J.~A.}, \bibinfo{author}{Fan, H.} \&
  \bibinfo{author}{Shaffer, J.~P.}
\newblock \bibinfo{title}{Collective state synthesis in an optical cavity using
  {R}ydberg atom dipole blockade}.
\newblock \emph{\bibinfo{journal}{Journal of Physics B: Atomic, Molecular and
  Optical Physics}} \textbf{\bibinfo{volume}{49}}, \bibinfo{pages}{064014}
  (\bibinfo{year}{2016}).
\newblock \urlprefix\url{http://stacks.iop.org/0953-4075/49/i=6/a=064014}.

\bibitem{kiffner2013}
\bibinfo{author}{Kiffner, M.}, \bibinfo{author}{Li, W.} \&
  \bibinfo{author}{Jaksch, D.}
\newblock \bibinfo{title}{Magnetic monopoles and synthetic spin-orbit coupling
  in {R}ydberg macrodimers}.
\newblock \emph{\bibinfo{journal}{Physical Review Letters}}
  \textbf{\bibinfo{volume}{110}}, \bibinfo{pages}{170402}
  (\bibinfo{year}{2013}).

\bibitem{Buchler2015}
\bibinfo{author}{Peter, D.} \emph{et~al.}
\newblock \bibinfo{title}{Topological bands with a chern number $c=2$ by
  dipolar exchange interactions}.
\newblock \emph{\bibinfo{journal}{Phys. Rev. A}} \textbf{\bibinfo{volume}{91}},
  \bibinfo{pages}{053617} (\bibinfo{year}{2015}).
\newblock \urlprefix\url{https://link.aps.org/doi/10.1103/PhysRevA.91.053617}.

\bibitem{bendkowsky2009}
\bibinfo{author}{Bendkowsky, V.} \emph{et~al.}
\newblock \bibinfo{title}{Observation of ultralong-range {R}ydberg molecules}.
\newblock \emph{\bibinfo{journal}{Nature}} \textbf{\bibinfo{volume}{458}},
  \bibinfo{pages}{1005--1008} (\bibinfo{year}{2009}).
\newblock \bibinfo{note}{\\{\bf This paper describes the first experimental
  observation of ultralong-range Rydberg molecules.}}

\bibitem{Tallant2012}
\bibinfo{author}{Tallant, J.}, \bibinfo{author}{Rittenhouse, S.~T.},
  \bibinfo{author}{Booth, D.}, \bibinfo{author}{Sadeghpour, H.~R.} \&
  \bibinfo{author}{Shaffer, J.~P.}
\newblock \bibinfo{title}{{Observation of Blueshifted Ultralong-Range
  ${\mathrm{Cs}}_{2}$ {R}ydberg Molecules}}.
\newblock \emph{\bibinfo{journal}{Physical Review Letters}}
  \textbf{\bibinfo{volume}{109}}, \bibinfo{pages}{173202}
  (\bibinfo{year}{2012}).
\newblock
  \urlprefix\url{http://link.aps.org/doi/10.1103/PhysRevLett.109.173202}.

\bibitem{Booth2015}
\bibinfo{author}{Booth, D.}, \bibinfo{author}{Rittenhouse, S.~T.},
  \bibinfo{author}{Yang, J.}, \bibinfo{author}{Sadeghpour, H.~R.} \&
  \bibinfo{author}{Shaffer, J.~P.}
\newblock \bibinfo{title}{{Production of trilobite {R}ydberg molecule dimers
  with kilo-Debye permanent electric dipole moments}}.
\newblock \emph{\bibinfo{journal}{Science}} \textbf{\bibinfo{volume}{348}},
  \bibinfo{pages}{99 -- 102} (\bibinfo{year}{2015}).
\newblock \urlprefix\url{http://science.sciencemag.org/content/348/6230/99}.
\newblock \bibinfo{note}{\\{\bf This paper describes the first experimental
  observation of trilobite molecules.}}

\bibitem{Bellos2013}
\bibinfo{author}{Bellos, M.~A.} \emph{et~al.}
\newblock \bibinfo{title}{Excitation of weakly bound molecules to trilobitelike
  {R}ydberg states}.
\newblock \emph{\bibinfo{journal}{Phys. Rev. Lett.}}
  \textbf{\bibinfo{volume}{111}}, \bibinfo{pages}{053001}
  (\bibinfo{year}{2013}).
\newblock
  \urlprefix\url{https://link.aps.org/doi/10.1103/PhysRevLett.111.053001}.

\bibitem{Anderson2014}
\bibinfo{author}{Anderson, D.~A.}, \bibinfo{author}{Miller, S.~A.} \&
  \bibinfo{author}{Raithel, G.}
\newblock \bibinfo{title}{Photoassociation of long-range {$nD$} {R}ydberg
  molecules}.
\newblock \emph{\bibinfo{journal}{Phys. Rev. Lett.}}
  \textbf{\bibinfo{volume}{112}}, \bibinfo{pages}{163201}
  (\bibinfo{year}{2014}).
\newblock
  \urlprefix\url{https://link.aps.org/doi/10.1103/PhysRevLett.112.163201}.
\newblock \bibinfo{note}{\\{\bf This paper was the first to predict the
  existence of spin mixed potentials in diatomic Rydberg molecules.}}

\bibitem{krupp2014}
\bibinfo{author}{Krupp, A.~T.} \emph{et~al.}
\newblock \bibinfo{title}{Alignment of $d$-state {R}ydberg molecules}.
\newblock \emph{\bibinfo{journal}{Phys. Rev. Lett.}}
  \textbf{\bibinfo{volume}{112}}, \bibinfo{pages}{143008}
  (\bibinfo{year}{2014}).
\newblock
  \urlprefix\url{https://link.aps.org/doi/10.1103/PhysRevLett.112.143008}.

\bibitem{Merkt2015}
\bibinfo{author}{Sa\ss{}mannshausen, H.}, \bibinfo{author}{Merkt, F.} \&
  \bibinfo{author}{Deiglmayr, J.}
\newblock \bibinfo{title}{Experimental characterization of singlet scattering
  channels in long-range {R}ydberg molecules}.
\newblock \emph{\bibinfo{journal}{Phys. Rev. Lett.}}
  \textbf{\bibinfo{volume}{114}}, \bibinfo{pages}{133201}
  (\bibinfo{year}{2015}).
\newblock
  \urlprefix\url{https://link.aps.org/doi/10.1103/PhysRevLett.114.133201}.

\bibitem{niederprum2016}
\bibinfo{author}{Niederpr{\"u}m, T.} \emph{et~al.}
\newblock \bibinfo{title}{Observation of pendular butterfly {R}ydberg
  molecules}.
\newblock \emph{\bibinfo{journal}{Nature Communications}}
  \textbf{\bibinfo{volume}{7}}, \bibinfo{pages}{12820} (\bibinfo{year}{2016}).
\newblock \urlprefix\url{http://dx.doi.org/10.1038/ncomms12820}.
\newblock \bibinfo{note}{\\{\bf This paper reported the first observation of
  p-wave interaction dominated ``butterfly'' molecules.}}

\bibitem{DeSalvo2015}
\bibinfo{author}{DeSalvo, B.~J.} \emph{et~al.}
\newblock \bibinfo{title}{Ultra-long-range {R}ydberg molecules in a divalent
  atomic system}.
\newblock \emph{\bibinfo{journal}{Phys. Rev. A}} \textbf{\bibinfo{volume}{92}},
  \bibinfo{pages}{031403} (\bibinfo{year}{2015}).
\newblock \urlprefix\url{https://link.aps.org/doi/10.1103/PhysRevA.92.031403}.

\bibitem{sadeghpour2013}
\bibinfo{author}{Sadeghpour, H.~R.} \& \bibinfo{author}{Rittenhouse, S.~T.}
\newblock \bibinfo{title}{How do ultralong-range homonuclear {R}ydberg
  molecules get their permanent dipole moments?}
\newblock \emph{\bibinfo{journal}{Molecular Physics}}
  \textbf{\bibinfo{volume}{111}}, \bibinfo{pages}{1902--1907}
  (\bibinfo{year}{2013}).
\newblock \urlprefix\url{https://doi.org/10.1080/00268976.2013.811555}.

\bibitem{rittenhouse2011}
\bibinfo{author}{Rittenhouse, S.~T.}, \bibinfo{author}{Mayle, M.},
  \bibinfo{author}{Schmelcher, P.} \& \bibinfo{author}{Sadeghpour, H.~R.}
\newblock \bibinfo{title}{Ultralong-range polyatomic {R}ydberg molecules formed
  by a polar perturber}.
\newblock \emph{\bibinfo{journal}{Journal of Physics B: Atomic, Molecular and
  Optical Physics}} \textbf{\bibinfo{volume}{44}}, \bibinfo{pages}{184005}
  (\bibinfo{year}{2011}).
\newblock
  \urlprefix\url{http://iopscience.iop.org/article/10.1088/0953-4075/44/18/184005}.

\bibitem{li2011}
\bibinfo{author}{Li, W.} \emph{et~al.}
\newblock \bibinfo{title}{A homonuclear molecule with a permanent electric
  dipole moment} \textbf{\bibinfo{volume}{334}}, \bibinfo{pages}{1110--1114}
  (\bibinfo{year}{2011}).
\newblock \urlprefix\url{http://science.sciencemag.org/content/334/6059/1110}.
\newblock \bibinfo{note}{\\{\bf The first observation of a permanent electric
  dipole moment in a homonuclear molecules due to the fractional mixture of
  ``trilobite'' electronic character was reported in this paper.}}

\bibitem{omont1977}
\bibinfo{author}{Omont, A.}
\newblock \bibinfo{title}{On the theory of collisions of atoms in {R}ydberg
  states with neutral particles}.
\newblock \emph{\bibinfo{journal}{Journal de Physique}}
  \textbf{\bibinfo{volume}{38}}, \bibinfo{pages}{1343--1359}
  (\bibinfo{year}{1977}).
\newblock
  \urlprefix\url{https://jphys.journaldephysique.org/articles/jphys/abs/1977/11/jphys_1977__38_11_1343_0/jphys_1977__38_11_1343_0.html}.

\bibitem{Bahrim2001b}
\bibinfo{author}{Bahrim, C.} \& \bibinfo{author}{Thumm, U.}
\newblock \bibinfo{title}{{Angle-differential and momentum-transfer cross
  sections for ${e}^{\ensuremath{-}}+\mathrm{Rb},$ Cs, and Fr collisions at low
  energies: ${}^{3}{F}^{o}$ shape resonances in
  ${\mathrm{Rb}}^{\ensuremath{-}},$ ${\mathrm{Cs}}^{\ensuremath{-}},$ and
  ${\mathrm{Fr}}^{\ensuremath{-}}$ ions}}.
\newblock \emph{\bibinfo{journal}{Phys. Rev. A}} \textbf{\bibinfo{volume}{64}},
  \bibinfo{pages}{022716} (\bibinfo{year}{2001}).
\newblock \urlprefix\url{https://link.aps.org/doi/10.1103/PhysRevA.64.022716}.

\bibitem{Bahrim2001a}
\bibinfo{author}{Bahrim, C.}, \bibinfo{author}{Thumm, U.} \&
  \bibinfo{author}{Fabrikant, I.~I.}
\newblock \bibinfo{title}{{$^3S_e$ and $^1S_e$ scattering lengths for e$^-$ +
  Rb, Cs and Fr collisions}}.
\newblock \emph{\bibinfo{journal}{Journal of Physics B}}
  \textbf{\bibinfo{volume}{34}}, \bibinfo{pages}{L195} (\bibinfo{year}{2001}).
\newblock \urlprefix\url{https://doi.org/10.1088/0953-4075/34/6/107}.

\bibitem{Bahrim2000}
\bibinfo{author}{Bahrim, C.} \& \bibinfo{author}{Thumm, U.}
\newblock \bibinfo{title}{{Low-lying ${}^{3}{P}^{o}$ and ${}^{3}{S}^{e}$ states
  of ${\mathrm{Rb}}^{\ensuremath{-}},{\mathrm{Cs}}^{\ensuremath{-}}$, and
  ${\mathrm{Fr}}^{\ensuremath{-}}$}}.
\newblock \emph{\bibinfo{journal}{Phys. Rev. A}} \textbf{\bibinfo{volume}{61}},
  \bibinfo{pages}{022722} (\bibinfo{year}{2000}).
\newblock \urlprefix\url{https://link.aps.org/doi/10.1103/PhysRevA.61.022722}.

\bibitem{hamilton2002}
\bibinfo{author}{Hamilton, E.~L.}, \bibinfo{author}{Greene, C.~H.} \&
  \bibinfo{author}{Sadeghpour, H.~R.}
\newblock \bibinfo{title}{Shape-resonance-induced long-range molecular
  {R}ydberg states}.
\newblock \emph{\bibinfo{journal}{Journal of Physics B: Atomic, Molecular and
  Optical Physics}} \textbf{\bibinfo{volume}{35}}, \bibinfo{pages}{L199}
  (\bibinfo{year}{2002}).
\newblock \urlprefix\url{https://doi.org/10.1088/0953-4075/35/10/102}.

\bibitem{Chibisov2002}
\bibinfo{author}{Chibisov, M.}, \bibinfo{author}{Khuskivadze, A.} \&
  \bibinfo{author}{Fabrikant, I.}
\newblock \bibinfo{title}{Energies and dipole moments of long-range molecular
  {R}ydberg states}.
\newblock \emph{\bibinfo{journal}{Journal of Physics B: Atomic, Molecular and
  Optical Physics}} \textbf{\bibinfo{volume}{35}}, \bibinfo{pages}{L193}
  (\bibinfo{year}{2002}).
\newblock \urlprefix\url{https://doi.org/10.1088/0953-4075/35/10/101}.

\bibitem{Raithel2014}
\bibinfo{author}{Anderson, D.~A.}, \bibinfo{author}{Miller, S.~A.} \&
  \bibinfo{author}{Raithel, G.}
\newblock \bibinfo{title}{Angular-momentum couplings in long-range
  ${\mathrm{rb}}_{2}$ {R}ydberg molecules}.
\newblock \emph{\bibinfo{journal}{Phys. Rev. A}} \textbf{\bibinfo{volume}{90}},
  \bibinfo{pages}{062518} (\bibinfo{year}{2014}).
\newblock \urlprefix\url{https://link.aps.org/doi/10.1103/PhysRevA.90.062518}.

\bibitem{Markson2016}
\bibinfo{author}{Markson, S.}, \bibinfo{author}{Rittenhouse, S.~T.},
  \bibinfo{author}{Schmidt, R.}, \bibinfo{author}{Shaffer, J.~P.} \&
  \bibinfo{author}{Sadeghpour, H.~R.}
\newblock \bibinfo{title}{Theory of ultralong-range {R}ydberg molecule
  formation incorporating spin-dependent relativistic effects: Cs(6s)–cs(np)
  as case study}.
\newblock \emph{\bibinfo{journal}{ChemPhysChem}} \textbf{\bibinfo{volume}{17}},
  \bibinfo{pages}{3683--3691} (\bibinfo{year}{2016}).
\newblock \urlprefix\url{http://dx.doi.org/10.1002/cphc.201600932}.

\bibitem{niederprum2016a}
\bibinfo{author}{Niederpr\"um, T.}, \bibinfo{author}{Thomas, O.},
  \bibinfo{author}{Eichert, T.} \& \bibinfo{author}{Ott, H.}
\newblock \bibinfo{title}{{R}ydberg molecule-induced remote spin flips}.
\newblock \emph{\bibinfo{journal}{Phys. Rev. Lett.}}
  \textbf{\bibinfo{volume}{117}}, \bibinfo{pages}{123002}
  (\bibinfo{year}{2016}).
\newblock
  \urlprefix\url{https://link.aps.org/doi/10.1103/PhysRevLett.117.123002}.

\bibitem{Kleinbach2017}
\bibinfo{author}{Kleinbach, K.~S.} \emph{et~al.}
\newblock \bibinfo{title}{Photoassociation of trilobite {R}ydberg molecules via
  resonant spin-orbit coupling}.
\newblock \emph{\bibinfo{journal}{Phys. Rev. Lett.}}
  \textbf{\bibinfo{volume}{118}}, \bibinfo{pages}{223001}
  (\bibinfo{year}{2017}).
\newblock
  \urlprefix\url{https://link.aps.org/doi/10.1103/PhysRevLett.118.223001}.

\bibitem{Bottcher2016}
\bibinfo{author}{B\"ottcher, F.} \emph{et~al.}
\newblock \bibinfo{title}{Observation of mixed singlet-triplet
  ${\mathrm{rb}}_{2}$ {R}ydberg molecules}.
\newblock \emph{\bibinfo{journal}{Phys. Rev. A}} \textbf{\bibinfo{volume}{93}},
  \bibinfo{pages}{032512} (\bibinfo{year}{2016}).
\newblock \urlprefix\url{https://link.aps.org/doi/10.1103/PhysRevA.93.032512}.

\bibitem{Eiles2017}
\bibinfo{author}{Eiles, M.~T.} \& \bibinfo{author}{Greene, C.~H.}
\newblock \bibinfo{title}{Hamiltonian for the inclusion of spin effects in
  long-range {R}ydberg molecules}.
\newblock \emph{\bibinfo{journal}{Phys. Rev. A}} \textbf{\bibinfo{volume}{95}},
  \bibinfo{pages}{042515} (\bibinfo{year}{2017}).
\newblock \urlprefix\url{https://link.aps.org/doi/10.1103/PhysRevA.95.042515}.

\bibitem{Pillet2010}
\bibinfo{author}{Comparat, D.} \& \bibinfo{author}{Pillet, P.}
\newblock \bibinfo{title}{Dipole blockade in a cold {R}ydberg atomic sample}.
\newblock \emph{\bibinfo{journal}{J. Opt. Soc. Am. B}}
  \textbf{\bibinfo{volume}{27}}, \bibinfo{pages}{A208} (\bibinfo{year}{2010}).
\newblock
  \urlprefix\url{http://josab.osa.org/abstract.cfm?URI=josab-27-6-A208}.

\bibitem{Gallagher2008}
\bibinfo{author}{Gallagher, T.~F.} \& \bibinfo{author}{Pillet, P.}
\newblock \bibinfo{title}{Dipole-dipole interactions of {R}ydberg atoms}.
\newblock In \bibinfo{editor}{et~al, A.} (ed.)
  \emph{\bibinfo{booktitle}{Advances in Atomic, Molecular, and Optical
  Physics}}, vol.~\bibinfo{volume}{56} of \emph{\bibinfo{series}{Advances In
  Atomic, Molecular, and Optical Physics}}, \bibinfo{pages}{161}
  (\bibinfo{publisher}{Academic Press}, \bibinfo{year}{2008}).
\newblock
  \urlprefix\url{http://www.sciencedirect.com/science/article/pii/S1049250X0800013X}.

\bibitem{Jones2012}
\bibinfo{author}{Vaillant, C.~L.}, \bibinfo{author}{Jones, M. P.~A.} \&
  \bibinfo{author}{Potvliege, R.~M.}
\newblock \bibinfo{title}{Long-range {R}ydberg-{R}ydberg interactions in
  {C}alcium, {S}trontium and {Y}tterbium}.
\newblock \emph{\bibinfo{journal}{J. Phys. B}} \textbf{\bibinfo{volume}{45}},
  \bibinfo{pages}{135004} (\bibinfo{year}{2012}).
\newblock \urlprefix\url{http://stacks.iop.org/0953-4075/45/i=13/a=135004}.

\bibitem{kiffner2012}
\bibinfo{author}{Kiffner, M.}, \bibinfo{author}{Park, H.}, \bibinfo{author}{Li,
  W.} \& \bibinfo{author}{Gallagher, T.~F.}
\newblock \bibinfo{title}{Dipole-dipole-coupled double-{R}ydberg molecules}.
\newblock \emph{\bibinfo{journal}{Phys. Rev. A}} \textbf{\bibinfo{volume}{86}},
  \bibinfo{pages}{031401} (\bibinfo{year}{2012}).
\newblock \urlprefix\url{https://link.aps.org/doi/10.1103/PhysRevA.86.031401}.

\bibitem{samboy2013}
\bibinfo{author}{Samboy, N.} \& \bibinfo{author}{C\^ot\'e, R.}
\newblock \bibinfo{title}{Rubidium {R}ydberg linear macrotrimers}.
\newblock \emph{\bibinfo{journal}{Phys. Rev. A}} \textbf{\bibinfo{volume}{87}},
  \bibinfo{pages}{032512} (\bibinfo{year}{2013}).
\newblock \urlprefix\url{https://link.aps.org/doi/10.1103/PhysRevA.87.032512}.

\bibitem{Nascimento2009}
\bibinfo{author}{Nascimento, V.~A.}, \bibinfo{author}{Caliri, L.~L.},
  \bibinfo{author}{Schwettmann, A.}, \bibinfo{author}{Shaffer, J.~P.} \&
  \bibinfo{author}{Marcassa, L.~G.}
\newblock \bibinfo{title}{{Electric field effects in the excitation of cold
  {R}ydberg-atom pairs}}.
\newblock \emph{\bibinfo{journal}{Phys. Rev. Lett.}}
  \textbf{\bibinfo{volume}{102}}, \bibinfo{pages}{213201}
  (\bibinfo{year}{2009}).
\newblock
  \urlprefix\url{https://link.aps.org/doi/10.1103/PhysRevLett.102.213201}.

\bibitem{Tallant2006}
\bibinfo{author}{Tallant, J.}, \bibinfo{author}{Overstreet, K.~R.},
  \bibinfo{author}{Schwettmann, A.} \& \bibinfo{author}{Shaffer, J.~P.}
\newblock \bibinfo{title}{Sub-doppler magneto-optical trap temperatures
  measured using {R}ydberg tagging}.
\newblock \emph{\bibinfo{journal}{Phys. Rev. A}} \textbf{\bibinfo{volume}{74}},
  \bibinfo{pages}{023410} (\bibinfo{year}{2006}).
\newblock \urlprefix\url{http://link.aps.org/doi/10.1103/PhysRevA.74.023410}.

\bibitem{Overstreet2007}
\bibinfo{author}{Overstreet, K.~R.}, \bibinfo{author}{Schwettmann, A.},
  \bibinfo{author}{Tallant, J.} \& \bibinfo{author}{Shaffer, J.~P.}
\newblock \bibinfo{title}{{Photoinitiated collisions between cold Cs {R}ydberg
  atoms}}.
\newblock \emph{\bibinfo{journal}{Phys. Rev. A}} \textbf{\bibinfo{volume}{76}},
  \bibinfo{pages}{011403} (\bibinfo{year}{2007}).
\newblock \urlprefix\url{https://link.aps.org/doi/10.1103/PhysRevA.76.011403}.

\bibitem{Browaeys2013}
\bibinfo{author}{B\'eguin, L.}, \bibinfo{author}{Vernier, A.},
  \bibinfo{author}{Chicireanu, R.}, \bibinfo{author}{Lahaye, T.} \&
  \bibinfo{author}{Browaeys, A.}
\newblock \bibinfo{title}{Direct measurement of the {V}an der {W}aals
  interaction between two {R}ydberg atoms}.
\newblock \emph{\bibinfo{journal}{Phys. Rev. Lett.}}
  \textbf{\bibinfo{volume}{110}}, \bibinfo{pages}{263201}
  (\bibinfo{year}{2013}).
\newblock
  \urlprefix\url{http://link.aps.org/doi/10.1103/PhysRevLett.110.263201}.

\bibitem{farooqi2003}
\bibinfo{author}{Farooqi, S.~M.} \emph{et~al.}
\newblock \bibinfo{title}{Long-range molecular resonances in a cold {R}ydberg
  gas}.
\newblock \emph{\bibinfo{journal}{Phys. Rev. Lett.}}
  \textbf{\bibinfo{volume}{91}}, \bibinfo{pages}{183002}
  (\bibinfo{year}{2003}).
\newblock
  \urlprefix\url{https://link.aps.org/doi/10.1103/PhysRevLett.91.183002}.

\bibitem{stanojevic2008}
\bibinfo{author}{Stanojevic, J.}, \bibinfo{author}{C\^ot\'e, R.},
  \bibinfo{author}{Tong, D.}, \bibinfo{author}{Eyler, E.~E.} \&
  \bibinfo{author}{Gould, P.~L.}
\newblock \bibinfo{title}{Long-range potentials and $(n\ensuremath{-}1)d+ns$
  molecular resonances in an ultracold {R}ydberg gas}.
\newblock \emph{\bibinfo{journal}{Phys. Rev. A}} \textbf{\bibinfo{volume}{78}},
  \bibinfo{pages}{052709} (\bibinfo{year}{2008}).
\newblock \urlprefix\url{https://link.aps.org/doi/10.1103/PhysRevA.78.052709}.

\bibitem{Yu2013}
\bibinfo{author}{Yu, Y.}, \bibinfo{author}{Park, H.} \&
  \bibinfo{author}{Gallagher, T.~F.}
\newblock \bibinfo{title}{Microwave transitions in pairs of {Rb} {R}ydberg
  atoms}.
\newblock \emph{\bibinfo{journal}{Phys. Rev. Lett.}}
  \textbf{\bibinfo{volume}{111}}, \bibinfo{pages}{173001}
  (\bibinfo{year}{2013}).
\newblock
  \urlprefix\url{http://link.aps.org/doi/10.1103/PhysRevLett.111.173001}.

\bibitem{Bendkowsky2010}
\bibinfo{author}{Bendkowsky, V.} \emph{et~al.}
\newblock \bibinfo{title}{{R}ydberg trimers and excited dimers bound by
  internal quantum reflection}.
\newblock \emph{\bibinfo{journal}{Phys. Rev. Lett.}}
  \textbf{\bibinfo{volume}{105}}, \bibinfo{pages}{163201}
  (\bibinfo{year}{2010}).
\newblock
  \urlprefix\url{https://link.aps.org/doi/10.1103/PhysRevLett.105.163201}.

\bibitem{Liu2006}
\bibinfo{author}{Liu, I. C.~H.} \& \bibinfo{author}{Rost, J.~M.}
\newblock \bibinfo{title}{Polyatomic molecules formed with a {R}ydberg atom in
  an ultracold environment}.
\newblock \emph{\bibinfo{journal}{The European Physical Journal D - Atomic,
  Molecular, Optical and Plasma Physics}} \textbf{\bibinfo{volume}{40}},
  \bibinfo{pages}{65--71} (\bibinfo{year}{2006}).
\newblock \urlprefix\url{https://doi.org/10.1140/epjd/e2006-00098-x}.

\bibitem{liu2009}
\bibinfo{author}{Liu, I. C.~H.}, \bibinfo{author}{Stanojevic, J.} \&
  \bibinfo{author}{Rost, J.~M.}
\newblock \bibinfo{title}{Ultra-long-range {R}ydberg trimers with a repulsive
  two-body interaction}.
\newblock \emph{\bibinfo{journal}{Phys. Rev. Lett.}}
  \textbf{\bibinfo{volume}{102}}, \bibinfo{pages}{173001}
  (\bibinfo{year}{2009}).
\newblock
  \urlprefix\url{https://link.aps.org/doi/10.1103/PhysRevLett.102.173001}.

\bibitem{Eiles2016}
\bibinfo{author}{Eiles, M.~T.}, \bibinfo{author}{P{\'e}rez-R{\'\i}os, J.},
  \bibinfo{author}{Robicheaux, F.} \& \bibinfo{author}{Greene, C.~H.}
\newblock \bibinfo{title}{Ultracold molecular {R}ydberg physics in a high
  density environment}.
\newblock \emph{\bibinfo{journal}{Journal of Physics B: Atomic, Molecular and
  Optical Physics}} \textbf{\bibinfo{volume}{49}}, \bibinfo{pages}{114005}
  (\bibinfo{year}{2016}).
\newblock \urlprefix\url{http://stacks.iop.org/0953-4075/49/i=11/a=114005}.

\bibitem{Rios2016}
\bibinfo{author}{P{\'e}rez-R{\'\i}os, J.}, \bibinfo{author}{Eiles, M.~T.} \&
  \bibinfo{author}{Greene, C.~H.}
\newblock \bibinfo{title}{Mapping trilobite state signatures in atomic
  hydrogen}.
\newblock \emph{\bibinfo{journal}{Journal of Physics B: Atomic, Molecular and
  Optical Physics}} \textbf{\bibinfo{volume}{49}}, \bibinfo{pages}{14LT01}
  (\bibinfo{year}{2016}).
\newblock \urlprefix\url{https://doi.org/10.1088/0953-4075/49/14/14LT01}.

\bibitem{gaj2014}
\bibinfo{author}{Gaj, A.} \emph{et~al.}
\newblock \bibinfo{title}{From molecular spectra to a density shift in dense
  {R}ydberg gases}.
\newblock \emph{\bibinfo{journal}{Nature Communications}}
  \textbf{\bibinfo{volume}{5}}, \bibinfo{pages}{4546} (\bibinfo{year}{2014}).
\newblock \urlprefix\url{http://dx.doi.org/10.1038/ncomms5546}.
\newblock \bibinfo{note}{\\{\bf The observation of larger oligomeric bound
  states with multiple ground state atoms bound to a single Rydberg atom was
  first reported in this paper.}}

\bibitem{Liebisch2016}
\bibinfo{author}{Liebisch, T.~C.} \emph{et~al.}
\newblock \bibinfo{title}{Controlling {R}ydberg atom excitations in dense
  background gases}.
\newblock \emph{\bibinfo{journal}{Journal of Physics B: Atomic, Molecular and
  Optical Physics}} \textbf{\bibinfo{volume}{49}}, \bibinfo{pages}{182001}
  (\bibinfo{year}{2016}).
\newblock \urlprefix\url{http://stacks.iop.org/0953-4075/49/i=18/a=182001}.

\bibitem{Fey2016}
\bibinfo{author}{Fey, C.}, \bibinfo{author}{Kurz, M.} \&
  \bibinfo{author}{Schmelcher, P.}
\newblock \bibinfo{title}{Stretching and bending dynamics in triatomic
  ultralong-range {R}ydberg molecules}.
\newblock \emph{\bibinfo{journal}{Phys. Rev. A}} \textbf{\bibinfo{volume}{94}},
  \bibinfo{pages}{012516} (\bibinfo{year}{2016}).
\newblock \urlprefix\url{https://link.aps.org/doi/10.1103/PhysRevA.94.012516}.

\bibitem{Fernandez2016}
\bibinfo{author}{Fern{\'a}ndez, J.~A.}, \bibinfo{author}{Schmelcher, P.} \&
  \bibinfo{author}{Gonz{\'a}lez-F{\'e}rez, R.}
\newblock \bibinfo{title}{Ultralong-range triatomic {R}ydberg molecules in an
  electric field}.
\newblock \emph{\bibinfo{journal}{Journal of Physics B: Atomic, Molecular and
  Optical Physics}} \textbf{\bibinfo{volume}{49}}, \bibinfo{pages}{124002}
  (\bibinfo{year}{2016}).
\newblock \urlprefix\url{http://stacks.iop.org/0953-4075/49/i=12/a=124002}.

\bibitem{Schmidt2016}
\bibinfo{author}{Schmidt, R.}, \bibinfo{author}{Sadeghpour, H.~R.} \&
  \bibinfo{author}{Demler, E.}
\newblock \bibinfo{title}{Mesoscopic {R}ydberg impurity in an atomic quantum
  gas}.
\newblock \emph{\bibinfo{journal}{Phys. Rev. Lett.}}
  \textbf{\bibinfo{volume}{116}}, \bibinfo{pages}{105302}
  (\bibinfo{year}{2016}).
\newblock
  \urlprefix\url{https://link.aps.org/doi/10.1103/PhysRevLett.116.105302}.

\bibitem{schlagmuller2016}
\bibinfo{author}{Schlagm\"uller, M.} \emph{et~al.}
\newblock \bibinfo{title}{Probing an electron scattering resonance using
  {R}ydberg molecules within a dense and ultracold gas}.
\newblock \emph{\bibinfo{journal}{Phys. Rev. Lett.}}
  \textbf{\bibinfo{volume}{116}}, \bibinfo{pages}{053001}
  (\bibinfo{year}{2016}).
\newblock
  \urlprefix\url{https://link.aps.org/doi/10.1103/PhysRevLett.116.053001}.

\bibitem{Camargo2016}
\bibinfo{author}{Camargo, F.} \emph{et~al.}
\newblock \bibinfo{title}{Lifetimes of ultra-long-range strontium {R}ydberg
  molecules}.
\newblock \emph{\bibinfo{journal}{Phys. Rev. A}} \textbf{\bibinfo{volume}{93}},
  \bibinfo{pages}{022702} (\bibinfo{year}{2016}).
\newblock \urlprefix\url{https://link.aps.org/doi/10.1103/PhysRevA.93.022702}.

\bibitem{Schmidt2017}
\bibinfo{author}{{Schmidt}, R.} \emph{et~al.}
\newblock \bibinfo{title}{{Theory of excitation of {R}ydberg polarons in an
  atomic quantum gas}}.
\newblock \bibinfo{howpublished}{Preprint at
  \url{https://arxiv.org/abs/1709.01838}, (2017)} (\bibinfo{year}{2017}).
\newblock \urlprefix\url{https://arxiv.org/abs/1709.01838}.

\bibitem{Camargo2017}
\bibinfo{author}{{Camargo}, F.} \emph{et~al.}
\newblock \bibinfo{title}{{Creation of {R}ydberg Polarons in a Bose Gas}}.
\newblock \bibinfo{howpublished}{Preprint at
  \url{https://arxiv.org/abs/1706.03717}, (2017)} (\bibinfo{year}{2017}).
\newblock \eprint{1706.03717}.

\bibitem{Butscher2011}
\bibinfo{author}{Butscher, B.} \emph{et~al.}
\newblock \bibinfo{title}{Lifetimes of ultralong-range {R}ydberg molecules in
  vibrational ground and excited states}.
\newblock \emph{\bibinfo{journal}{Journal of Physics B: Atomic, Molecular and
  Optical Physics}} \textbf{\bibinfo{volume}{44}}, \bibinfo{pages}{184004}
  (\bibinfo{year}{2011}).
\newblock \urlprefix\url{http://stacks.iop.org/0953-4075/44/i=18/a=184004}.

\bibitem{SChlagmuller2016b}
\bibinfo{author}{Schlagm\"uller, M.} \emph{et~al.}
\newblock \bibinfo{title}{Ultracold chemical reactions of a single {R}ydberg
  atom in a dense gas}.
\newblock \emph{\bibinfo{journal}{Phys. Rev. X}} \textbf{\bibinfo{volume}{6}},
  \bibinfo{pages}{031020} (\bibinfo{year}{2016}).
\newblock \urlprefix\url{https://link.aps.org/doi/10.1103/PhysRevX.6.031020}.

\bibitem{Sassmannshausen2016}
\bibinfo{author}{Sa{\ss}mannshausen, H.}, \bibinfo{author}{Deiglmayr, J.} \&
  \bibinfo{author}{Merkt, F.}
\newblock \bibinfo{title}{Long-range {R}ydberg molecules, {R}ydberg macrodimers
  and {R}ydberg aggregates in an ultracold cs gas}.
\newblock \emph{\bibinfo{journal}{The European Physical Journal Special
  Topics}} \textbf{\bibinfo{volume}{225}}, \bibinfo{pages}{2891--2918}
  (\bibinfo{year}{2016}).
\newblock \urlprefix\url{https://doi.org/10.1140/epjst/e2016-60124-9}.

\bibitem{Whalen2017}
\bibinfo{author}{Whalen, J.~D.} \emph{et~al.}
\newblock \bibinfo{title}{Lifetimes of ultralong-range strontium {R}ydberg
  molecules in a dense bose-einstein condensate}.
\newblock \emph{\bibinfo{journal}{Phys. Rev. A}} \textbf{\bibinfo{volume}{96}},
  \bibinfo{pages}{042702} (\bibinfo{year}{2017}).
\newblock \urlprefix\url{https://link.aps.org/doi/10.1103/PhysRevA.96.042702}.

\bibitem{Merkt2015b}
\bibinfo{author}{Sa\ss{}mannshausen, H.}, \bibinfo{author}{Merkt, F.} \&
  \bibinfo{author}{Deiglmayr, J.}
\newblock \bibinfo{title}{Pulsed excitation of {R}ydberg-atom-pair states in an
  ultracold cs gas}.
\newblock \emph{\bibinfo{journal}{Phys. Rev. A}} \textbf{\bibinfo{volume}{92}},
  \bibinfo{pages}{032505} (\bibinfo{year}{2015}).
\newblock \urlprefix\url{https://link.aps.org/doi/10.1103/PhysRevA.92.032505}.

\bibitem{rittenhouse2010}
\bibinfo{author}{Rittenhouse, S.~T.} \& \bibinfo{author}{Sadeghpour, H.~R.}
\newblock \bibinfo{title}{Ultracold giant polyatomic {R}ydberg molecules:
  Coherent control of molecular orientation}.
\newblock \emph{\bibinfo{journal}{Phys. Rev. Lett.}}
  \textbf{\bibinfo{volume}{104}}, \bibinfo{pages}{243002}
  (\bibinfo{year}{2010}).
\newblock
  \urlprefix\url{https://link.aps.org/doi/10.1103/PhysRevLett.104.243002}.
\newblock \bibinfo{note}{\\{\bf The existence of yet be observed Rydberg
  molecules consisting of an atom in a highly excited Rydberg state and a
  ground-state polar molecule were first predicted here.}}

\bibitem{mayle2012}
\bibinfo{author}{Mayle, M.}, \bibinfo{author}{Rittenhouse, S.~T.},
  \bibinfo{author}{Schmelcher, P.} \& \bibinfo{author}{Sadeghpour, H.~R.}
\newblock \bibinfo{title}{Electric field control in ultralong-range triatomic
  polar {R}ydberg molecules}.
\newblock \emph{\bibinfo{journal}{Phys. Rev. A}} \textbf{\bibinfo{volume}{85}},
  \bibinfo{pages}{052511} (\bibinfo{year}{2012}).
\newblock \urlprefix\url{https://link.aps.org/doi/10.1103/PhysRevA.85.052511}.

\bibitem{Gonzalez2015}
\bibinfo{author}{Gonz{\'a}lez-F{\'e}rez, R.}, \bibinfo{author}{Sadeghpour,
  H.~R.} \& \bibinfo{author}{Schmelcher, P.}
\newblock \bibinfo{title}{Rotational hybridization, and control of alignment
  and orientation in triatomic ultralong-range {R}ydberg molecules}.
\newblock \emph{\bibinfo{journal}{New Journal of Physics}}
  \textbf{\bibinfo{volume}{17}}, \bibinfo{pages}{013021}
  (\bibinfo{year}{2015}).
\newblock \urlprefix\url{http://stacks.iop.org/1367-2630/17/i=1/a=013021}.

\bibitem{Kuznetsova2011}
\bibinfo{author}{Kuznetsova, E.}, \bibinfo{author}{Rittenhouse, S.~T.},
  \bibinfo{author}{Sadeghpour, H.~R.} \& \bibinfo{author}{Yelin, S.~F.}
\newblock \bibinfo{title}{{R}ydberg atom mediated polar molecule interactions:
  a tool for molecular-state conditional quantum gates and individual
  addressability}.
\newblock \emph{\bibinfo{journal}{Phys. Chem. Chem. Phys.}}
  \textbf{\bibinfo{volume}{13}}, \bibinfo{pages}{17115--17121}
  (\bibinfo{year}{2011}).
\newblock \urlprefix\url{http://dx.doi.org/10.1039/C1CP21476D}.

\bibitem{Kuznetsova2016}
\bibinfo{author}{Kuznetsova, E.}, \bibinfo{author}{Rittenhouse, S.~T.},
  \bibinfo{author}{Sadeghpour, H.~R.} \& \bibinfo{author}{Yelin, S.~F.}
\newblock \bibinfo{title}{{R}ydberg-atom-mediated nondestructive readout of
  collective rotational states in polar-molecule arrays}.
\newblock \emph{\bibinfo{journal}{Phys. Rev. A}} \textbf{\bibinfo{volume}{94}},
  \bibinfo{pages}{032325} (\bibinfo{year}{2016}).
\newblock \urlprefix\url{https://link.aps.org/doi/10.1103/PhysRevA.94.032325}.

\bibitem{Balewski2013}
\bibinfo{author}{Balewski, J.~B.} \emph{et~al.}
\newblock \bibinfo{title}{Coupling a single electron to a bose--einstein
  condensate}.
\newblock \emph{\bibinfo{journal}{Nature}} \textbf{\bibinfo{volume}{502}},
  \bibinfo{pages}{664--667} (\bibinfo{year}{2013}).
\newblock \urlprefix\url{http://dx.doi.org/10.1038/nature12592}.

\bibitem{Ott2015}
\bibinfo{author}{Manthey, T.}, \bibinfo{author}{Niederpr{\"u}m, T.},
  \bibinfo{author}{Thomas, O.} \& \bibinfo{author}{Ott, H.}
\newblock \bibinfo{title}{Dynamically probing ultracold lattice gases via
  {R}ydberg molecules}.
\newblock \emph{\bibinfo{journal}{New Journal of Physics}}
  \textbf{\bibinfo{volume}{17}}, \bibinfo{pages}{103024}
  (\bibinfo{year}{2015}).
\newblock \urlprefix\url{http://stacks.iop.org/1367-2630/17/i=10/a=103024}.

\bibitem{Luukko2017}
\bibinfo{author}{Luukko, P. J.~J.} \& \bibinfo{author}{Rost, J.-M.}
\newblock \bibinfo{title}{Polyatomic trilobite {R}ydberg molecules in a dense
  random gas}.
\newblock \emph{\bibinfo{journal}{Phys. Rev. Lett.}}
  \textbf{\bibinfo{volume}{119}}, \bibinfo{pages}{203001}
  (\bibinfo{year}{2017}).
\newblock
  \urlprefix\url{https://link.aps.org/doi/10.1103/PhysRevLett.119.203001}.

\bibitem{Maxwell2013}
\bibinfo{author}{Maxwell, D.} \emph{et~al.}
\newblock \bibinfo{title}{Storage and control of optical photons using
  {R}ydberg polaritons}.
\newblock \emph{\bibinfo{journal}{Phys. Rev. Lett.}}
  \textbf{\bibinfo{volume}{110}}, \bibinfo{pages}{103001}
  (\bibinfo{year}{2013}).
\newblock
  \urlprefix\url{https://link.aps.org/doi/10.1103/PhysRevLett.110.103001}.

\bibitem{Thompson2017}
\bibinfo{author}{Thompson, J.~D.} \emph{et~al.}
\newblock \bibinfo{title}{Symmetry-protected collisions between strongly
  interacting photons}.
\newblock \emph{\bibinfo{journal}{Nature}} \textbf{\bibinfo{volume}{542}},
  \bibinfo{pages}{206--209} (\bibinfo{year}{2017}).
\newblock \urlprefix\url{http://dx.doi.org/10.1038/nature20823}.

\bibitem{Mirgorodskiy2017}
\bibinfo{author}{Mirgorodskiy, I.} \emph{et~al.}
\newblock \bibinfo{title}{Electromagnetically induced transparency of
  ultra-long-range {R}ydberg molecules}.
\newblock \emph{\bibinfo{journal}{Phys. Rev. A}} \textbf{\bibinfo{volume}{96}},
  \bibinfo{pages}{011402} (\bibinfo{year}{2017}).
\newblock \urlprefix\url{https://link.aps.org/doi/10.1103/PhysRevA.96.011402}.

\bibitem{Sandor2017}
\bibinfo{author}{S\'andor, N.}, \bibinfo{author}{Gonz\'alez-F\'erez, R.},
  \bibinfo{author}{Julienne, P.~S.} \& \bibinfo{author}{Pupillo, G.}
\newblock \bibinfo{title}{{R}ydberg optical feshbach resonances in cold gases}.
\newblock \emph{\bibinfo{journal}{Phys. Rev. A}} \textbf{\bibinfo{volume}{96}},
  \bibinfo{pages}{032719} (\bibinfo{year}{2017}).
\newblock \urlprefix\url{https://link.aps.org/doi/10.1103/PhysRevA.96.032719}.

\bibitem{Thomas2017}
\bibinfo{author}{{Thomas}, O.}, \bibinfo{author}{{Lippe}, C.},
  \bibinfo{author}{{Eichert}, T.} \& \bibinfo{author}{{Ott}, H.}
\newblock \bibinfo{title}{{Experimental realization of a {R}ydberg optical
  Feshbach resonance in a quantum many-body system}}.
\newblock \bibinfo{howpublished}{Preprint at
  \url{https://arxiv.org/abs/1712.05263}, (2017)} (\bibinfo{year}{2017}).

\bibitem{Butscher2010}
\bibinfo{author}{Butscher, B.} \emph{et~al.}
\newblock \bibinfo{title}{Atom-molecule coherence for ultralong-range {R}ydberg
  dimers}.
\newblock \emph{\bibinfo{journal}{Nature Physics}}
  \textbf{\bibinfo{volume}{6}}, \bibinfo{pages}{970--974}
  (\bibinfo{year}{2010}).

\bibitem{Kirrander2013}
\bibinfo{author}{Kirrander, A.} \emph{et~al.}
\newblock \bibinfo{title}{{Approach to form long-range ion-pair molecules in an
  ultracold Rb gas}}.
\newblock \emph{\bibinfo{journal}{Phys. Rev. A}} \textbf{\bibinfo{volume}{87}},
  \bibinfo{pages}{031402} (\bibinfo{year}{2013}).
\newblock \urlprefix\url{https://link.aps.org/doi/10.1103/PhysRevA.87.031402}.

\bibitem{Markson2016a}
\bibinfo{author}{Markson, S.} \& \bibinfo{author}{Sadeghpour, H.~R.}
\newblock \bibinfo{title}{{A model for charge transfer in ultracold {R}ydberg
  ground-state atomic collisions}}.
\newblock \emph{\bibinfo{journal}{Journal of Physics B: Atomic, Molecular and
  Optical Physics}} \textbf{\bibinfo{volume}{49}}, \bibinfo{pages}{114006}
  (\bibinfo{year}{2016}).
\newblock \urlprefix\url{http://stacks.iop.org/0953-4075/49/i=11/a=114006}.

\bibitem{Wilczek1986}
\bibinfo{author}{Moody, J.}, \bibinfo{author}{Shapere, A.} \&
  \bibinfo{author}{Wilczek, F.}
\newblock \bibinfo{title}{Realizations of magnetic-monopole gauge fields:
  Diatoms and spin precession}.
\newblock \emph{\bibinfo{journal}{Phys. Rev. Lett.}}
  \textbf{\bibinfo{volume}{56}}, \bibinfo{pages}{893} (\bibinfo{year}{1986}).
\newblock \urlprefix\url{http://link.aps.org/doi/10.1103/PhysRevLett.56.893}.

\bibitem{Cederbaum1989}
\bibinfo{author}{Pacher, T.}, \bibinfo{author}{Mead, C.~A.},
  \bibinfo{author}{Cederbaum, L.~S.} \& \bibinfo{author}{Köppel, H.}
\newblock \bibinfo{title}{Gauge theory and quasidiabatic states in molecular
  physics}.
\newblock \emph{\bibinfo{journal}{The Journal of Chemical Physics}}
  \textbf{\bibinfo{volume}{91}}, \bibinfo{pages}{7057--7062}
  (\bibinfo{year}{1989}).
\newblock \urlprefix\url{http://dx.doi.org/10.1063/1.457323}.

\bibitem{Baer2002}
\bibinfo{author}{Baer, M.}
\newblock \bibinfo{title}{Introduction to the theory of electronic
  non-adiabatic coupling terms in molecular systems}.
\newblock \emph{\bibinfo{journal}{Phys. Rep.}} \textbf{\bibinfo{volume}{358}},
  \bibinfo{pages}{75} (\bibinfo{year}{2002}).
\newblock
  \urlprefix\url{http://www.sciencedirect.com/science/article/pii/S0370157301000527}.

\bibitem{khuskivadze2002}
\bibinfo{author}{Khuskivadze, A.~A.}, \bibinfo{author}{Chibisov, M.~I.} \&
  \bibinfo{author}{Fabrikant, I.~I.}
\newblock \bibinfo{title}{{Adiabatic energy levels and electric dipole moments
  of {R}ydberg states of ${\mathrm{Rb}}_{2}$ and ${\mathrm{Cs}}_{2}$ dimers}}.
\newblock \emph{\bibinfo{journal}{Phys. Rev. A}} \textbf{\bibinfo{volume}{66}},
  \bibinfo{pages}{042709} (\bibinfo{year}{2002}).
\newblock \urlprefix\url{https://link.aps.org/doi/10.1103/PhysRevA.66.042709}.

\end{thebibliography}
\end{document}